\UseRawInputEncoding
\documentclass[%
 reprint,
floatfix,
superscriptaddress,
showpacs,preprintnumbers,
 amsmath,amssymb,
 aps,
pra,
]{revtex4-1}
\usepackage{xcolor}
\usepackage{times}
\usepackage{amssymb}
\usepackage[notrig]{physics}
\usepackage{graphicx}% Include figure files
\usepackage{dcolumn}% Align table columns on decimal point
\usepackage{bm}% bold math
\usepackage{epstopdf}
\usepackage{hyperref}% add hypertext capabilities

\begin{document}
\title{\textbf{Strange metal from incoherent bosons}}

\author{Anurag Banerjee}
\affiliation{Institut de Physique Th\'eorique, Universit\'{e} Paris-Saclay, CEA, CNRS, F-91191 Gif-sur-Yvette, France}

\author{Maxence Grandadam} 
\affiliation{Institut de Physique Th\'eorique, Universit\'{e} Paris-Saclay, CEA, CNRS, F-91191 Gif-sur-Yvette, France}

\author{Hermann Freire}
\affiliation{Instituto de F\'{i}­sica, Universidade Federal de Goi\'{a}s, 74.001-970 Goi\^{a}nia-GO, Brazil}

\author{Catherine P\'epin}
\affiliation{Institut de Physique Th\'eorique, Universit\'{e} Paris-Saclay, CEA, CNRS, F-91191 Gif-sur-Yvette, France}

% Please give the surname of the lead author for the running footer
\begin{abstract}
The breakdown of the celebrated Fermi liquid theory in the strange metal phase is the central enigma of correlated quantum matter. Motivated by recent experiments reporting short-lived carriers, along with the ubiquitous observations of modulated excitations in the phase diagram of cuprates, we propose a model for this phase. We introduce bosons emerging from the remnants of a pair density wave as additional current carriers in the strange metal phase. These bosonic excitations are finite momentum Cooper pairs and thus carry twice the electronic charge, and its net spin can either be zero or one arising from the two spin-$1/2$ electrons. We show that such a model can capture the famous linear relationship of resistivity with temperature and manifests the Drude form of ac-conductivity with a Planckian dissipation rate. Furthermore, such bosons are incoherent and hence do not contribute to the Hall conductivity. The bosons emerging from the electron pairs of spin-triplet symmetry also reproduces the recently observed linear in field magnetoresistance~[P. Giraldo-Gallo \textit{et al.}, Science \textbf{361}, 479 (2018); J. Ayres \textit{et al.}, \emph{arXiv:} 2012.01208 (2020)].
\end{abstract}

\maketitle
\section{Introduction}
\label{introduction}
One common thread among diverse strongly correlated materials is the emergence of an anomalous metallic state upon destroying superconductivity~\cite{gurvitch1987resistivity,SMTBG20,StewartRMP,rost11_Ruth}. This emergent state is referred to as bad metal when the conventional quasiparticle concept becomes invalid at high temperatures~\cite{gurvitch1987resistivity,EmeryVJ:1995es,hussey2004universality}, whereas in strange metal, such anomaly extends down to very low $T$~\cite{cooper2009anomalous}. Particularly in cuprates, over a vast temperature-doping region, the resistivity shows a linear-in-$T$ dependence from low temperature up to the melting point of the material~\cite{gurvitch1987resistivity, Waszczak_Exp1,hussey2008phenomenology,AlloulPRB11,AlloulPRL2007}, thus manifesting both bad and strange metallic characteristics. This behavior is recently associated with the `Planckian' dissipation rate, $\hbar\tau^{-1}\sim k_{B}T$, which is the maximal dissipation rate allowed by the laws of quantum mechanics~\cite{legros2019universal,bruin2013similarity,zaanen2019planckian,cha2020linear}. Interestingly, for frequencies lower than such dissipation rate, i.e., $\omega<\tau^{-1}$, the optical conductivity remarkably follows the classic Drude form~\cite{van2003quantum, zaanen2019planckian}, in addition to showing a linear-in-$T$ resistivity. 

Recently, the strange metal (SM) is gaining impetus with the observation of mysterious incoherent carriers
in the optimally-doped and overdoped cuprates~\cite{hashimoto2009crossover,chen2019incoherent}. Over the region where the dc-resistivity is most linear, there is a
significant reduction of the Hall carriers~\cite{putzke2019reduced,PRBGreven20}, suggesting short-lived carriers responsible for the transport. Furthermore, at high
magnetic fields, the magnetoresistance also displays a linear in field evolution in hole-doped cuprates~\cite{giraldo2018scale, Hussey2020}
which is further confirmed in other compounds~\cite{sarkar2019correlation,HusseyFe_MR,hayes2016scaling}. 
Such incoherent conductivity is insensitive to the magnetic field's orientation, 
again implying a vanishing Hall conductivity~\cite{Hussey2020}.
Thus, the mysterious SM phase acquires another element: On the one hand, it shows
linear-in-$T$ resistivity with the optical conductivity following the classic Drude form,
and with an additional incoherent transport component insensitive to the orientation
of the magnetic field. On the other hand, the experimental result since
the dawn of the cuprates~\cite{clayhold1989hall,Hall_Manko_92,barivsic19} exhibits a second transport time $\hbar\tau_{H}^{-1}\sim T^{2}$ 
which controls the cotangent of the Hall angle\footnote{Defined as $\cot\theta_H=\sigma_{xx}/\sigma_{xy}$ 
where $\sigma_{xx}$ and $\sigma_{xy}$ are, respectively, the longitudinal and the Hall conductivities.} over the whole phase diagram. 
A consistent theory for strange metal must reconcile all these
unusual behaviors, which still remains a fundamental challenge in condensed matter physics. 

Early attempts to demystify the strange metal phase rely on the rationale that the fermionic excitations are primarily responsible for its odd transport properties. 
These theories capture some basic features of the SM phase; for example, the marginal Fermi liquid theory~\cite{VarmaMFL},
among others~\cite{PAnderson91,HiddenFL, Coleman96}, can heuristically describe the
temperature dependence of longitudinal conductivity and the Hall angle~\cite{Varma_MRL_hall}.
More recently, the Hall transport time, $\tau_{H}$, are satisfactorily
described by the presence of quasielectrons with an anisotropic transport
time around the Fermi surface~\cite{RossMcKenziePRL11,RossMcKenzie12,hussey2013generic,hussey2003coherent}. 
It is also highlighted that interactions can improve nesting near the hot-spots in the spin fermion model~\cite{chubukov_adv_phys,tsvelik2017ladder}, 
which can lead to $T$-linear resistivity with a broad Drude component~\cite{classen19}. Furthermore, 
such a model can capture the $T^2$-dependence of the cotangent of the Hall angle\cite{rice2017umklapp}.
However, most theories presently encounter difficulties in accounting for the linear-in-$T$ resistivity and the corresponding 
Planckian limit of the scattering rate. Furthermore, the Drude form of the optical conductivity,
along with the recent report of incoherent non-orbital contribution to transport~\cite{Hussey2020} remains to be addressed.
Given that situation, a regime of very strong coupling, obtained by either holographic techniques~\cite{hartnoll2018holographic,faulkner2010strange,hartnoll2015theory,BlakeHallAngle15,ZaanenRPB_Holo,amoretti2019universal}
or other transport methods~\cite{wu2018candidate,patel2018magnetotransport,ErezBerg20,Paul13,McKenzieDMFT00,huang19strange} have been invoked to account for some of these observed properties. 

To address this challenging problem, an intuitive phenomenological model is imperative. Motivated by the recent
discovery of incoherent carriers~\cite{hashimoto2009crossover,chen2019incoherent} along with the ubiquitous 
observations of spatially modulating patterns~\cite{Hamidian16, Wang18}
in the phase diagram of cuprates, we introduce a strange metal model that provides a significant perspective shift.
We propose bosons emerging from the spatially undulating electron-electron pairs as additional current carriers in the strange metal phase. 
These bosonic excitations are remnants of a pair density wave (PDW) state and consequently carry twice the charge of an electron,
including a finite wave-vector linked to its vestigial periodicity.
The net spin of the boson due to its constituent spin-1/2 electrons, consequently, can either have a spin-singlet
or spin-triplet symmetry.
Therefore, the fermionic quasielectrons are not the sole charge carriers in this phase. 
We show that that the charged bosons become diffusive and incoherent as they interact with the underlying fermions.
Within this scenario, the quasielectrons around
the Fermi surface naturally account for the observed coherent transport
in the material, since they react to the magnetic field according
to the Hall lifetime $\tau_{H}$.  In contrast, the bosons provide a natural explanation
for the incoherent transport reported recently, which we discuss below.

\section{The model \label{subsec:The-model}}

We propose a model consisting of quasielectrons scattering off each other via hydrodynamic fluctuations as well as charge-two bosons. The bosons originate from pairs of high-energy electrons, which interact with the low-energy quasielectrons, with strength, $g_I$, and with themselves with strength, $g_b$. With the application of an external magnetic field, the corresponding gauge-invariant Hamiltonian becomes 
\begin{align}
\hat{\mathcal{H}} & =\sum_{\mathbf{k},\alpha}c_{\mathbf{k},\alpha}^{\dagger}\left[\frac{\left(\mathbf{k}-e\mathbf{A}\right)^{2}}{2m}-\epsilon_{F}\right]c_{\mathbf{k},\alpha}+V_{e-e}
\nonumber\\
&+\sum_{\mathbf{q}}b_{\mathbf{q}}^{\dagger}\left[{\frac{1}{4}( \mathbf{Q}_0+\delta\mathbf{q}-2e\mathbf{A})^{2}}+\mu_0\right]b_{\mathbf{q}} \nonumber \\ & -\frac{1}{2}\sum_{\mathbf{k},\alpha,\alpha'}c_{\mathbf{k},\alpha}^{\dagger}(\vec{\sigma}_{\alpha\alpha'}.\mathbf{H})c_{\mathbf{k},\alpha'} +g_{b}\sum_{\mathbf{q},\mathbf{p},\mathbf{k}}b_{\mathbf{k}}^{\dagger}b_{\mathbf{k}+\mathbf{q}}b_{\mathbf{p}-\mathbf{q}}^{\dagger}b_{\mathbf{p}} \nonumber\\
 & +g_{I}\sum_{\mathbf{k},\mathbf{q},\alpha}\left[b_{\mathbf{q}}^{\dagger}c_{\mathbf{k},\alpha}c_{-\mathbf{k}+\mathbf{q},\pm\alpha}+h.c.\right]
 ,\nonumber\\
\label{model}
\end{align}
where $c_{\mathbf{k,\alpha}}^{\dagger}$ is the creation operator
for conduction electrons, $\alpha$ is the spin projection of the electrons, and
$b_{\mathbf{q}}^{\dagger}$ is the creation operator for charge-two bosons.  Our idea is that the finite-momentum Cooper pairing fluctuations, with wave vector $\mathbf{Q}_0$, are forming at intermediate temperatures under strong coupling. Once the finite-momentum Cooper pair fluctuations are formed, gauge invariance imposes the vector potential is associated to $\mathbf{Q}_0$. We have used $\mathbf{q}=\mathbf{Q}_0+\delta\mathbf{q}$ such that $\delta\mathbf{q} \ll \mathbf{Q}_0$. The quantities $e$ and $m$ are, respectively, the elementary charge and the quasielectron mass, whereas $\mathbf{A}$ is the vector potential associated with the magnetic field given by $\mathbf{H}=\nabla\times\mathbf{A}$. The quantities $\epsilon_{F}$ and $\mu_0$ denote, respectively, the chemical potential of the electrons and the bare bosonic mass term. The next term refers to the coupling of the electron spins to the Zeeman field, where $\vec{\sigma}_{\alpha\alpha'}$ are the Pauli matrices. The term $V_{e-e}$ represents the interactions between the electrons and the environment that can consist of other types of hydrodynamic modes or impurities. Finally, the
last two terms in Eq.~(\ref{model}) are, respectively, the boson-boson interaction and the fermion-boson interaction. In the interaction term that contains $g_{I}$, we allow for the possibility of the bosons to be either spin-0 or spin-1. 

\section{Results \label{subsec:Longitudinal-conductivity}}
\begin{figure}[h!]
\centering
\includegraphics[width=0.95\linewidth]{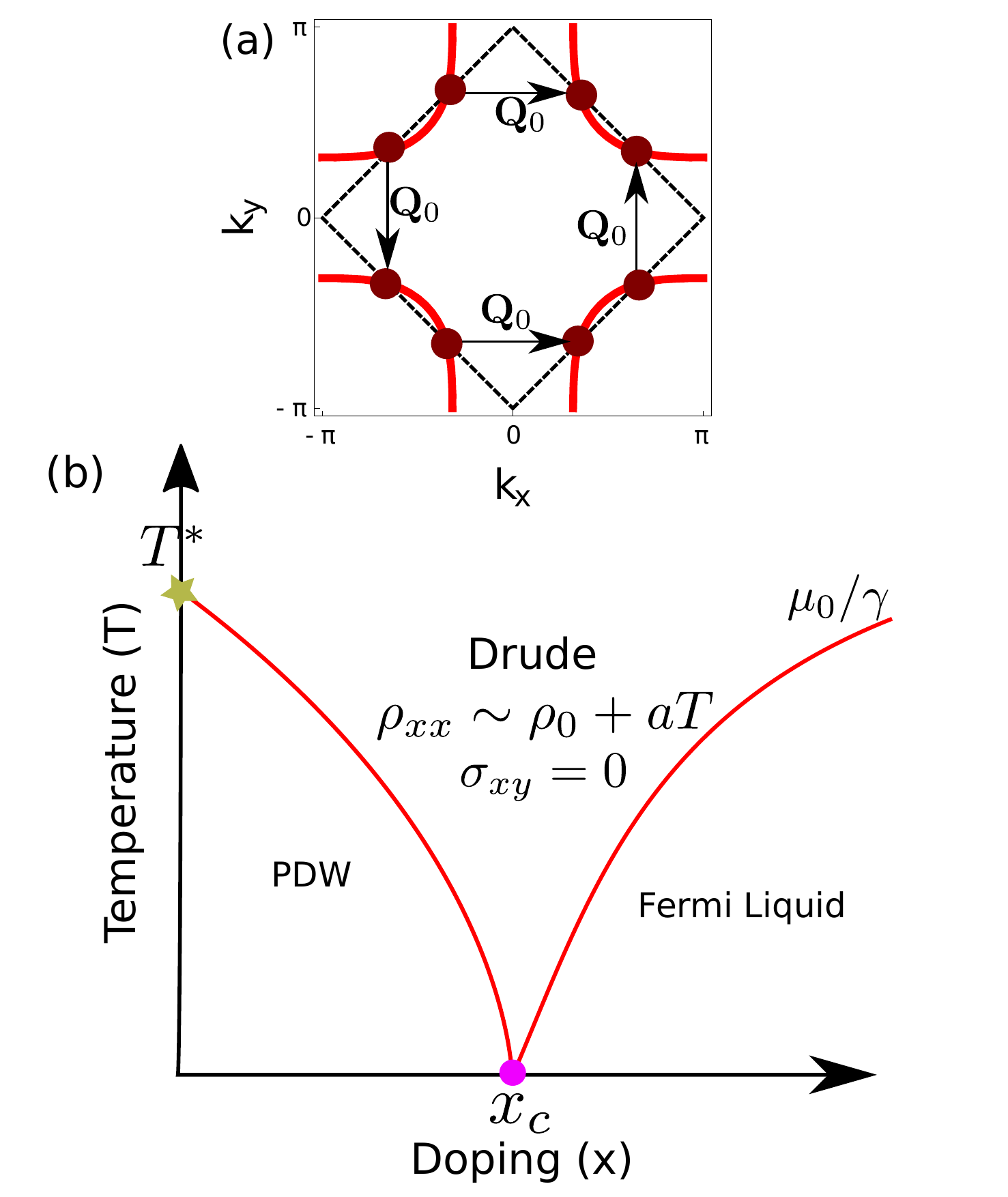}
\caption{\textbf{(a)} Shows the Fermi surface observed in overdoped cuprates with the hot-spots denoted by brown circles. \textbf{(b)} Displays a skeleton phase diagram in the temperature-doping plane. The $T^{*}$ sets the PDW energy scale in the system, which vanishes at the QCP, $x_c$. For larger dopings, the Fermi liquid behavior is established. The strange metal phase is expected to reside in the quantum critical fan in between these two regions. The energy scale that separates the two distinct regions is $\mu_0/\gamma$, where $\mu_0$ is the doping-dependent bare bosonic mass term, and the $\gamma$ is the Landau damping coefficient of the diffusive bosons. Depending on the size of $\mathbf{Q}_0$, a slightly different phase diagram (details in Appendix~\ref{App:FBVertex}) is possible, in which the theory is valid up to a low temperature but not down to $T=0$.}
\label{Fig:schm} 
\end{figure}

\begin{figure}[h!]
\includegraphics[width=0.45\textwidth]{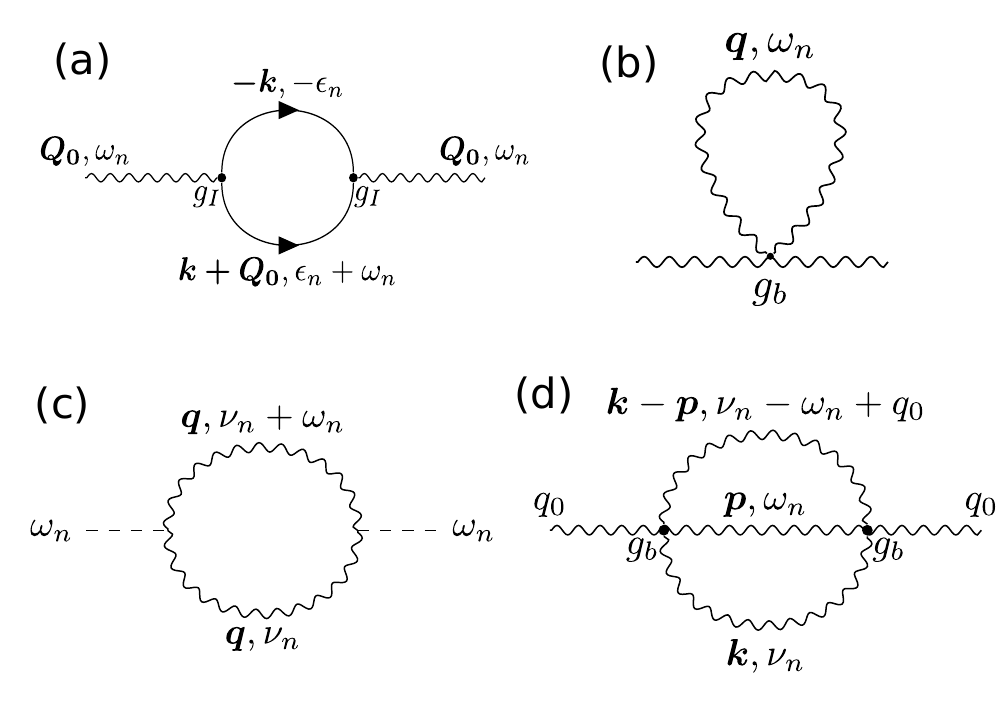}
\caption{Feynman diagrams corresponding to the transport properties
and the interactions of the bosons among themselves and with the fermions
of the model defined in Eq.~(\ref{model}): (a), (b) and (d) represent contributions to the bosonic self-energy in the present theory, whereas (c) stands for the diagram associated with the current-current correlation function.}
\label{Fig:diag} 
\end{figure}

We study the electromagnetic response of the system within the Kubo formalism, considering not only the electronic but also the bosonic response to the electromagnetic field. The Feynman diagrams contributing to the charge transport properties and the self-energy corrections are presented in Fig.~(\ref{Fig:diag}). The bosons originate from the strongly coupled electrons, and hence the effective mass of these bosons is expected to be smaller than the strongly correlated fermionic quasiparticles. Consequently, the bosonic carriers dominate the longitudinal conductivity in our model. However, since the bosonic excitations have a particle-hole symmetry, the Hall conductivity vanishes for these bosons, as shown in our subsequent analysis. Therefore, the Hall conductivity is dominated by the fermionic quasiparticles giving a $T^2$-dependence of the cotangent of the Hall angle~\cite{RossMcKenziePRL11,RossMcKenzie12}. 

Naturally, charged bosons have a markedly different behavior from fermions. At low temperatures, fermions scatter around the Fermi surface, and scattering with finite wavevectors affects only small regions of the Fermi surface, creating a transport anisotropy commonly referred to as ``hot spots'' and ``cold spots''~\cite{RiceHotSpot}. The hot-spots are shown by the circles in the Fig.~(\ref{Fig:schm}a). Such fermions participate both in the transport and in the Drude weight~\cite{van2003quantum,lobo2011optical,hussey2003coherent}. On the other hand, bosons do not have a Fermi surface and, consequently, they scatter uniformly through other species in the sample. Therefore, the bosonic pathway of charge transport is protected against short circuit of hot regions by the cold ones, unlike the fermionic counterpart~\cite{RiceHotSpot}. 

A momentum relaxation mechanism is necessary to obtain a steady current flow upon applying an external electric field~\cite{pal2012resistivity}. The
incoherent bosons are dynamical fluctuations with a particle-hole symmetry
and therefore are in a hydrodynamic regime. Consequently, incoherent bosons have
a lifetime linked to its transport time. Moreover, these finite momentum
bosons described by ${b_{\mathbf{q}}^{\dagger}= c_{\mathbf{k}}^{\dagger}c_{-\mathbf{k}+\mathbf{q}}^{\dagger}}$
are made of pairs of electrons on the Fermi surface of Fig.~(\ref{Fig:schm}a). Thus, these bosons itinerant on a lattice are
akin to phonons, paving the way for multiple scattering mechanisms to decay the current~\cite{ziman}.
Consequently, the two species (fermions and the finite-momentum bosons) contribute to the transport,
and both terms must be included to obtain the total optical sum rule. The present study is devoted to a careful analysis
of the finite momentum bosonic contribution to the transport. In contrast, zero momentum bosons require other
mechanisms that can break the Galilean symmetry, as in the case of paraconductivity~\cite{BoyackPRB_bosons,larkin2005theory}. 

When the coupling between the bosons is stronger than the damping coefficient, our key findings are encapsulated in the phase diagram of Fig.~(\ref{Fig:schm}c). Above a threshold temperature $T>\mu_0/\gamma$, we find a linear-in-$T$ resistivity and a vanishing Hall conductance (where $\gamma$ is the Landau-damping coefficient of the diffusive bosons to be defined shortly). Here, $\mu_0$ is the doping-dependent bare mass of the boson, which vanishes at the quantum critical point or a critical phase. We emphasize that our phenomenological study cannot distinguish between a quantum critical point and quantum critical phase as observed in Ref.~\cite{husseycupratescriticality}. Furthermore, when $T>\mu_0/\gamma$, the incoherent bosons contribute to the Drude-like conductivity with a scattering rate reminiscent of Planckian dissipation~\cite{zaanen04_Nature,zaanen2019planckian,bruin2013similarity,legros2019universal}. On the other hand, when the temperature is below $T<\mu_0/\gamma$, the traditional Fermi-liquid behavior is established due to the additional presence of a fermionic pathway~\cite{RossMcKenzie12,hussey2013generic}. The ratio of the bare bosonic mass, $\mu_0$, to the damping strength of the bosons determines the crossover from the strange metallic to conventional metal regime, as exhibited in Fig.~(\ref{Fig:schm}c).

\subsection{Boson scattering via the fermions}
We consider the scattering process of bosons from electrons as the predominant one. Evaluating the diagram on Fig.~(\ref{Fig:diag}a) (detailed calculations are given in Appendix~\ref{App:Scattering}), 
we note that such polarization bubble is proportional to $g_{I}^{2}$ and produces a Landau damping term.
This distinctive feature is typical of a charge-two boson with finite momentum, which couples to electrons in the same way as a pair-density-wave (PDW).  
After integrating out the electronic degrees of freedom, the bosonic propagator reads
\begin{align}
\mathcal{D}^{-1}({\mathbf{q},i\omega_{n}}) & =\gamma\left|\omega_{n}\right|+{\mathbf{q}^{2}}+\mu(T).
\label{boson_prop}
\end{align}
Here, $\omega_{n}$ is the Matsubara frequency, where the Landau-damping constant is given by ${\gamma= g^2_I \mathcal{N}(\epsilon_{F}) /( 2 \pi\sqrt{(2 k_F Q_0)^2-Q_0^4})}$, where $\mathcal{N}(\epsilon_{F})$ is density of states at the Fermi energy, $k_F$ is the corresponding Fermi momentum and $\mu(T)$ is the bosonic ``mass-term'' at finite temperatures. This form of the bosonic Green's function is valid for all the frequencies below $\omega_c \approx k_F Q_0$. 

Next, we comment on the effects of fermion-boson vertex corrections in the present theory. Recent studies of the antiferromagnetic QCP~\cite{Metlitski10b,lee18NFL} in two spatial dimensions obtained that the vertex corrections yield logarithmic divergences. Such divergences renormalize the dynamical exponent at the QCP from the initial $z=2$ towards a smaller value.  In the present study of bosons with a finite wave-vector, two different situations can emerge. In the first one, the bosons cannot generate hotspots if the wave-vector $\mathbf{Q}_0$ is either too small or too large to connect distinct parts of the Fermi surface. In this scenario, the Landau damping remains unchanged, whereas the vertex corrections become irrelevant (for details, see Appendix~\ref{App:RobLD}). Hence, it gives both a $T$-linear resistivity and a broad Drude component extending to zero temperature. In the second scenario, if the bosons create hotspots by connecting different parts of the Fermi surface, the vertex corrections should become relevant and effectively change the dynamical exponent $z$ near the QCP. Nevertheless, this renormalization of $z$ is expected to occur only at very low temperatures near the QCP~\cite{gerlach17}. Above this temperature, other damping sources (See Appendix~\ref{App:FBVertex}) can regularize the vertex corrections and recover the linear-in-$T$ behavior with the Drude form of ac-conductivity over a broad temperature range.

From Eq.~(\ref{boson_prop}), it becomes clear that the bosons are diffusive near the critical point (or critical phase) where the bare mass of the boson vanishes. Moreover, the form factor of the electron pairs does not have a qualitative influence on the diffusive behavior of the bosons. We have checked numerically, e.g., that a $d$-wave form factor for the electron pairs also leads to such Landau damping term, albeit with a different coefficient. We show below that the bosonic propagator of Eq.~(\ref{boson_prop}) can contribute to the incoherent part of the resistivity that was recently reported in Ref.~\cite{chen2019incoherent,Hussey2020}. 

\subsection{Kubo formula for the conductivity}

Since the charge-two boson directly couples to the electromagnetic field, the main bosonic contribution to the longitudinal resistivity is given within the Kubo formula by the diagram in Fig.~(\ref{Fig:diag}c) (see Appendix~\ref{App:Kubo}, for
detailed evaluation of this diagram). The leading-order contribution to the conductivity is given by 
\begin{align}
\sigma_{ij}(\omega) =\frac{T}{\omega_{n}}\sum_{\varepsilon_{n}} &\int dx\int dx^\prime \left\lbrace -\delta_{i j} \delta(\mathbf{x}-\mathbf{x^{\prime}}) \mathcal{D}(\varepsilon_{n},x,x') \right. \nonumber \\
& \left. +\hat{v}_{i} \mathcal{D}(\varepsilon_{n},x,x') \hat{v}_{j} \mathcal{D}(\varepsilon_{n}+\omega_{n},x',x) \right\rbrace,\label{eq:3}
\end{align}
where the analytical continuation $i\omega_{n}\rightarrow\omega+i\delta$
needs to be performed, the indices $i,j$ refer to the spatial directional, $\hat{v}_{x}=\left(-i\partial_{x}-iH\partial_{k_{y}}\right)$ and
$\hat{v}_{y}=\left(-i\partial_{y}+iH\partial_{k_{x}}\right)$ are the velocity kernels. The longitudinal conductivity (independent of the magnetic field $H$) is then given by 
\begin{align}
\sigma^{(0)}_{xx}  (\omega_n)=\frac{T}{\omega_{n}} \sum_{\varepsilon_{n}} \frac{1}{L}\sum_{\mathbf{q}} \left[ Q_{0}^{2} \mathcal{D}(\varepsilon_{n},\mathbf{q}) \mathcal{D}(\varepsilon_{n}+\omega_{n},\mathbf{q})  \right. \nonumber \\
 \left. + \mathcal{D}(\varepsilon_n,\mathbf{q}) \right] .\label{eq:4}
\end{align}
Note that since the bosons have a finite momentum, the velocity kernels in Eq.~(\ref{eq:3})
become proportional to $\mathbf{Q}_{0}$, which result in a prefactor for the above integral. Thus, performing the corresponding integration,
we find in the first regime, i.e., $\gamma T \ll \mu$, that the optical conductivity becomes
\begin{align}
\sigma_{xx}\left(\omega\right) & =\frac{\sigma_{0}^{b} \tau}{\left( 1- i\frac{\gamma \omega}{2 \mu} \right)},\label{eq:5}
\end{align}
with $\sigma_{0}^{b}=Q_0^2/(2 \pi^2 \gamma)$. Strikingly,
Eq.~(\ref{eq:5}) is reminiscent of a Drude conductivity, with the scattering rate given by $\hbar\tau^{-1}=(2\mu/\gamma)$. 

However, in the second regime, i.e., $\gamma T \gg \mu$, the optical conductivity does not exhibit the traditional Drude form
\begin{equation}
\sigma(\omega)=\frac{Q_0^2 \mu }{12\pi^2 \gamma^2 T^2} \left( 1- i\frac{\gamma \omega}{2 \mu} \right).
\label{sigma0_2}
\end{equation}
We will show in the next section that this latter regime (non-Drude-like) is never obtained if the coupling strength between the bosons is larger
than the Landau-damping parameter.

\subsection{Renormalization of the bosonic ``mass-term'' }
\begin{figure}[h!]
\includegraphics[width=0.5\textwidth]{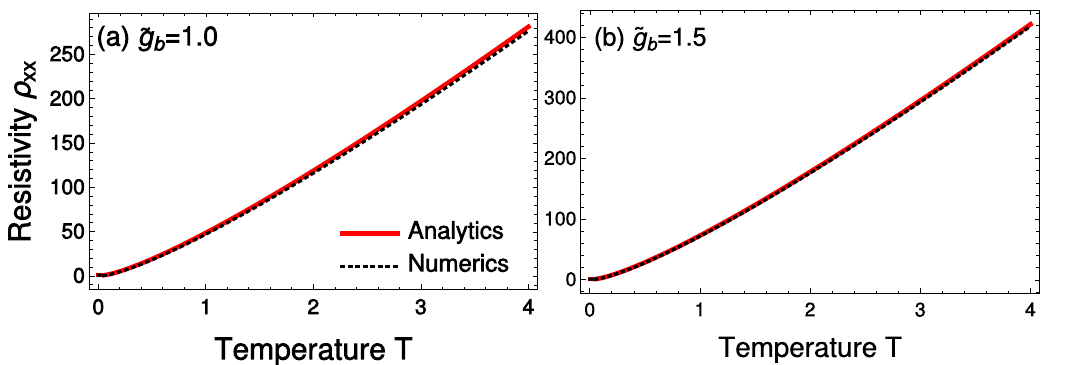}
\caption{Displays the linear-in-$T$ evolution of resistivity obtained from the analysis of the model. In all the plots, we set the Landau-damping constant equal to $\gamma=1.0$ and the temperature independent mass term $\mu_0=0.05$. The temperature is given in units of $\mu_0/\gamma$. The following physical constants are set to unity: $\hbar=1$, $k_B=1$, and $e=1$. Besides, we choose also the input parameters \textbf{(a)} $\tilde{g}_b=1.0 $ and \textbf{(b)} $\tilde{g}_b=1.5$. Above $T>\mu_0/\gamma$ the linear-in-$T$ behavior sets in. We also compare both the numerical and the analytical expressions in these plots which are in good agreement with each other.}
\label{Fig:resistivty}  
\end{figure}

In order to figure out the temperature dependence of the static resistivity, 
we evaluate the renormalization of the bosonic ``mass-term'' due to its scattering with strength $g_b$. 
This is given by the diagram in Fig.~(\ref{Fig:diag}b), which is proportional to the number of bosons, $N_b=T\sum_{\nu_{n}}\sum_{\mathbf{q}} \mathcal{D}(\nu_{n},\mathbf{q})$.
The bosonic mass term of the Eq.~(\ref{boson_prop}) is renormalized by
\begin{equation}
\mu=\mu_{0}+ g_{b} N_b,
\end{equation}
where $\mu_0$ is the bare mass-term. The leading-order correction to the mass-term evaluates to (for details refer to Appendix~\ref{App:Nb})
\begin{align}
\mu = \begin{cases} 
\mu_0+ \tilde{g}_b T\log\left( \frac{\gamma T}{\mu_0} \right) & \mbox{for   } \gamma T \gg \mu_0, \\
\mu_0 & \mbox{for   } \gamma T \ll \mu_0,
\end{cases} 
\label{eq:6}
\end{align}
where we have defined, $\tilde{g}_b=g_b/(4 \pi)$. Therefore, for an intermediate to strong coupling regime, i.e., $\tilde{g}_b\ge\gamma$, we will always have $\gamma T \ll \mu$. For this reason, the second regime displaying the non-Drude form of the optical conductivity is not attained if the coupling is stronger than the damping. In the main text, we mainly focus on the $\tilde{g}_b\ge\gamma$ regime. The possibility of the other theoretical limits are explored in Appendix~\ref{App:Static}.  Thus, plugging the temperature-dependence of the bosonic ``mass-term'' calculated in Eq.~(\ref{eq:6}) back into Eq.~(\ref{eq:5}), the static $\omega \rightarrow 0$ becomes 
\begin{align}
\rho_{xx}(T) = \begin{cases}
\frac{4 \pi^2 \mu_0}{Q_0^2} +\frac{ 4 \pi^2 \tilde{g}_b }{Q_0^2}T \log\left(\frac{\gamma T}{\mu_0}\right) & \mbox{for   } \gamma T \gg \mu_0,\\
\frac{4 \pi^2 \mu_0}{Q_0^2} & \mbox{for   } \gamma T \ll \mu_0.
\end{cases} 
\label{eq:8}
\end{align}
Therefore, up to logarithmic corrections, we obtain a linear-in-$T$ regime for the resistivity
when $ \gamma T\geq \mu_{0}$, with no saturation at large temperatures, thus capturing the bad metal
regime. The first term is a $T$-independent contribution that can vanish when the bosons become critical. 
Consequently, the $T$-linear resistivity can extend up to zero temperature, thereby achieving
the strange metal regime in the optimally doped and overdoped cuprates.  
To further confirm our analytical results, we perform numerical integration to obtain the static resistivity as
a function of temperature. The Fig.~(\ref{Fig:resistivty}a) shows a clear linear-in-$T$ behavior
of resistivity for the following parameter choices, $\gamma=1.0$, $Q_0=\pi/2$, $\mu_0=0.05$ and $\tilde{g}_b=1.0$. As can be
seen, there is a very good match between the numerical and the approximate analytical behavior.
Similarly, in Fig.~(\ref{Fig:resistivty}b) we show the results for a larger interaction parameter $\tilde{g}_b=1.5$, 
which again displays a linear dependence with temperature, albeit with a
different slope (for details, see Appendix~\ref{App:Kubo}).

\begin{figure}[h!]
\includegraphics[width=0.5\textwidth]{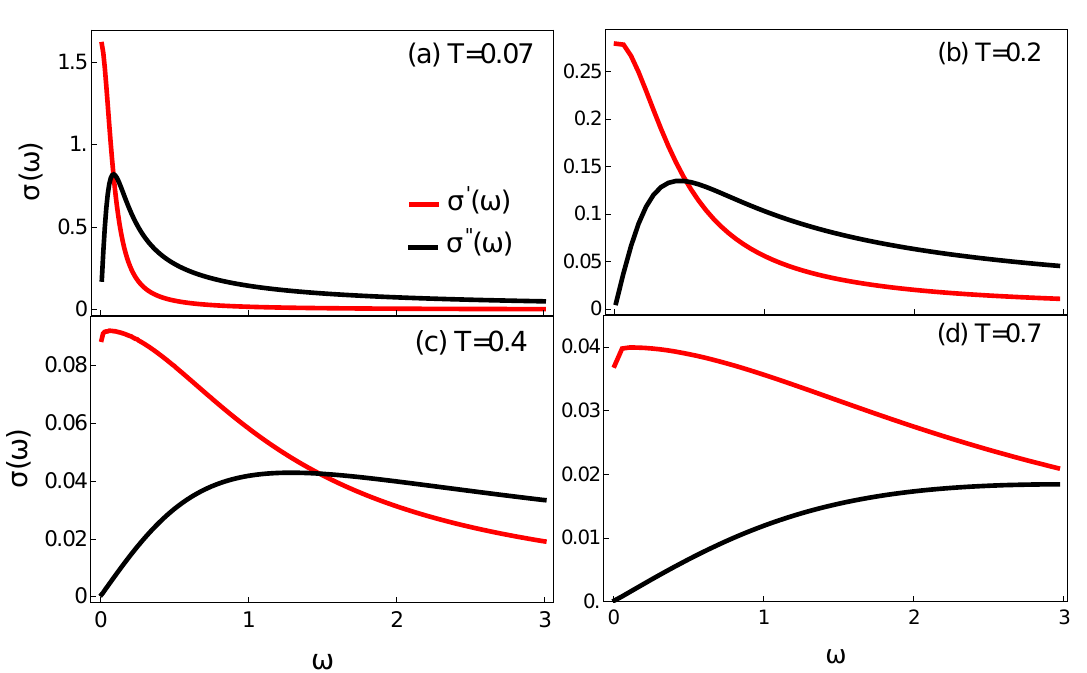}
\caption{Shows real and imaginary parts of the optical conductivity $\sigma(\omega)=\sigma^{\prime}(\omega)+i \sigma^{\prime \prime}(\omega)$ for the following parameter choices: $\gamma=1.0$, $\tilde{g}_b=1.0$ and $\mu_0=0.05.$ \textbf{(a)} $T=0.07$, \textbf{(b)} $T=0.2$, \textbf{(c)} $T=0.4$ and \textbf{(d)} $T=0.7$. The temperature is given in units of $\mu_0/\gamma$. The following physical constants are set to unity: $\hbar=1$, $k_B=1$, and $e=1$. Here, $\sigma(\omega)$ shows the traditional Drude form with the width of real part increasing with temperature. Thus, the linear-in-$T$ resistivity from the incoherent bosons contributes to a Drude-like response at finite frequencies.}
\label{Fig:Drude} 
\end{figure}

Moreover, our calculations reveal that the incoherent transport due to the charged bosons contributes to the Drude-like response at finite frequencies. Furthermore, in this regime, the transport momentum relaxation rate, ${\tau^{-1} \sim k_B T/\hbar}$, scales linearly with temperature up to logarithmic corrections. In overdoped cuprates,  line-shapes of the optical conductivity as a function of frequency remarkably follow such classic Drude form~\cite{van2003quantum}. Furthermore, a close relationship between the scattering rates of the charge carriers and linear-in-$T$ behavior is established across several families of overdoped cuprates~\cite{legros2019universal,bruin2013similarity}.  In Fig.~(\ref{Fig:resistivty}a), we present the full optical conductivity as a function of frequency $\omega$ for the following parameter choices, $\gamma=1.0$, $Q_0=\pi/2$, $\mu_0=0.05$ and $\tilde{g}_b=1.0$. The real part of the optical conductivity exhibits a sharp peak at a low temperature, $T=0.07$, similar to ac-conductivity experiments in the ``good strange metal'' regime. The peak broadens progressively as the temperature increases to $T=0.7$, as presented in Fig.~(\ref{Fig:Drude}b) to Fig.~(\ref{Fig:Drude}d). We obtain from Eq.~(\ref{eq:8}) a longitudinal conductivity that varies as $T^{-1}$ (up to logarithmic corrections), which participates in a Drude-like response at finite frequency. This is an astonishing outcome of our theory. The Planckian dissipation within the holographic framework appears to be a highly generic feature of the dense many-body entangled quantum matter~\cite{hartnoll2018holographic,faulkner2010strange,BlakeHallAngle15,ZaanenRPB_Holo,hartnoll2015theory,amoretti2019universal}. However, within such a holographic duality approach, it is still undecided whether a fixed point can produce incoherent transport contributing to the Drude conductivity. Our straightforward model illustrates such Drude behavior with a Planckian dissipation rate.

\subsection{Higher order terms in the self-energy}
\label{subsec:MMcoupling}

We now turn to the next-to-leading order correction regarding the ``mass-term'' renormalization,  namely, the rainbow
diagram represented in Fig.~(\ref{Fig:diag}d).
In addition, the imaginary part from this diagram renormalizes the Landau-damping
constant $\gamma$ in Eq.~(\ref{boson_prop}). The corresponding polarization bubble reads
\begin{align}
\Pi_{2}(q_{0})= g_{b}^{2}T^{2}\sum_{\nu_{n},\omega_{n}}&\sum_{\mathbf{p},\mathbf{q}} \mathcal{D}(\nu_n-\omega_n+q_0,\mathbf{-p+k})\nonumber\\
& \times \mathcal{D}(\nu_n,\mathbf{k}) \mathcal{D}(\omega_n,\mathbf{p}),
\end{align}
where $q_0$ is the incoming frequency that is assumed to be a small parameter during the evaluation. The renormalization of the $\mu$ and $\gamma$ to the second order for $\gamma T \gg \mu_0$ is given by (details presented in Appendix~\ref{App:Pi2})
\begin{eqnarray}
\mu &\approx& \mu_0 + \frac{g_b}{4 \pi } \log\left(\frac{\gamma T}{\mu_0}\right)+ \frac{2 c_1 \gamma \lambda }{\pi^2 \log^2\left( \gamma T/\mu_0 \right)}, \\
\tilde{\gamma}&\approx&\gamma+\frac{c_1 \gamma }{ \pi \log^2\left( \gamma T/\mu_0 \right)},
\end{eqnarray}
where $\lambda=\text{min}[\mu_0,\gamma T]$. Now we take the limit ${\gamma T / \mu_0 \gg 1}$ and find that the second-order terms are negligible.
Next, evaluating the same quantities for $\gamma T \ll \mu_0$, we get
\begin{eqnarray}
\mu &\approx& \mu_0 + \frac{c_2 \lambda (\gamma T)^4 }{2 \pi^6 \gamma \mu_0^4}, \\
\tilde{\gamma}&\approx&\gamma+\frac{c_2 (\gamma T)^3}{4 \pi^5 \gamma \mu_0^4}.
\end{eqnarray}
If we assume ${\gamma T / \mu_0 \ll 1}$, the second order contributions then become negligible. The constants $c_1=0.323$ and $c_2=0.284$ are evaluated by employing numerical techniques. Therefore, in both regimes, the higher-order terms are small compared to the first-order ones and, therefore, we can safely ignore their effects from now on in our analysis. Moreover, the vertex-correction diagram at second-order in the coupling $g_b$ is of the same order of magnitude as the bosonic self-energy $\Pi_{2}$ and, as demonstrated in Appendix~\ref{App:Vertex}, we can also ignore this contribution in our analysis.

\subsection{Hall conductivity}
We begin the discussion on the effect of magnetic field on the charged bosons with the Hall conductivity.
The term linear-in-$H$ term in Eq.~(\ref{eq:3}) leads to the following expression for the Hall conductivity, which is given by
\begin{align}
\sigma_{xy}^{\left(1\right)} ={\frac{1}{\omega_{n}}\text{Im } \bigg\{T\sum_{\varepsilon_{n}} \frac{1}{L} \sum_{\mathbf{q}}{iH}[q_{x}\mathcal{D}({\varepsilon_{n},\mathbf{q}})\partial_{q_{y}} \mathcal{D}({\varepsilon_{n}+\omega_{n},\mathbf{q})}}\nonumber \\
  -\partial_{q_{y}} \mathcal{D}({\varepsilon_{n},\mathbf{q}})\,{q_{x}} \mathcal{D}({\varepsilon_{n}+\omega_{n},\mathbf{q}})]\bigg\}.\label{eq:9}
\end{align}
Evaluating this term with Eq.~(\ref{boson_prop}), we obtain that it naturally vanishes (details can be found in Appendix~\ref{App:Hall}).
This result is not surprising, since the bosons are incoherent and the theory has a particle-hole symmetry. This can be confirmed by noting
that the bosonic propagator in Eq.~(\ref{boson_prop}) is symmetric under $\omega \rightarrow - \omega$ transformation. The fact that
diffusive bosons do not participate in the Hall number could explain
the recent studies where the number
of Hall carriers is seen to gradually decrease, as the doping is reduced
from the overdoped region to the underdoped regime~\cite{putzke2019reduced,PRBGreven20}. Similarly, vanishing Hall conductivity
is reported in the normal state of the stripe-ordered cuprates~\cite{shi2019magnetic,li2019tuning}, and in two-dimensional superconducting thin-films~\cite{Armitage18,breznay2017particle}.
The emergence of particle-hole symmetry of the charged incoherent bosons in this study also implicates a tendency towards the vanishing Hall conductivity.

\subsection{Second-moment of conductivity}
The contribution quadratic in $H$ of the conductance in Eq.~(\ref{eq:3}) writes 
\begin{align}
\sigma^{(2)}_{xx}  =-\frac{H^2}{\omega_{n}}\text{Im }\bigg\{T\sum_{\varepsilon_{n}}\sum_{\mathbf{q}}\left[\partial_{q_{x}}D({\varepsilon_{n}+\omega_{n},\mathbf{q}}) \partial_{q_{x}}D({\varepsilon_{n},\mathbf{q}})\right]\bigg\}.\label{eq:10}
\end{align}
This orbital contribution from Eq.~(\ref{eq:10}) is calculated in Appendix~\ref{App:Sigma2} and, in the regime $\gamma T \ll \mu$, it reads
\begin{align}
\sigma^{(2)}_{xx}=\frac{ 8 \gamma^2 Q_0^2 T^2 H^2}{ 5 \pi^2 \mu^5}.
\label{sigma2}
\end{align}
Again, we emphasize that the second regime, $\gamma T \gg \mu$, is never realized when the interaction between the electrons is stronger than the Landau damping coefficient. For completeness, we provide the corresponding expressions for the same in SI. Armed with the expression for $\sigma_{xx}^{(0)}$, $\sigma^{(1)}_{xx}$, and $\sigma^{(2)}_{xx}$, we proceed to evaluate the magnetic field dependence of the magnetoresistance.

\subsection{Magnetoresistance} 
For a system with vanishing Hall conductivity $\sigma_{xy}$, the magnetoresistance is evaluated (details provided in Appendix~\ref{App:MR}) through 
\begin{equation}
\frac{\Delta\rho_{xx}}{\rho_{xx}(0)}=\frac{\rho_{xx} (H)-\rho_{xx} (0)}{\rho_{xx} (0)}=\frac{\sigma_{xx}(0)-\sigma_{xx}(H)}{\sigma_{xx}(H)},
\end{equation}
where $\sigma_{xx}(0)$ denotes the conductance measured at
zero magnetic field. The longitudinal conductivity, however, has contributions 
from both $\sigma^{(0)}_{xx}$ and $\sigma^{(2)}_{xx}$. In order to proceed, the mass renormalization due to the Zeeman field needs to be evaluated.
Two cases then arise due to the symmetry of the spins of the electron pairs.

\subsubsection{Spin-zero case}
First, let us consider that the diffusive bosons have spin-zero, i.e., the spins of the electron pairs have the symmetry of a singlet. The
Zeeman coupling to the spin of the electrons (diagram in Fig.~(\ref{Fig:diag}a)) renormalizes the bosonic mass term. The resulting renormalization is independent of the magnetic field $H$ and is given by
\begin{equation}
\mu=\mu_0+\mu_T,
\end{equation}
where $\mu_T=\tilde{g}_b T \log(\gamma T/\mu_0)$ (details provided in Appendix~\ref{App:Singlet}).
On the other hand, since the orbital contribution Eq.~(\ref{eq:10}) gives a term in quadratic in $H$, it leads to a
$H^2$ dependence of the MR (evaluated in detail in Appendix~\ref{App:Singlet}). The regimes are then determined by the
maximum among $\mu_0$ and $\mu_T$. The MR is given by
\begin{align}
\frac{\Delta \rho_{xx}}{\rho_{xx}(0)}=\frac{\kappa}{\beta} H^2,
\end{align}
where $\kappa/\beta\equiv-32 \gamma^2 T^2/( 5 \mu^4)$, with $\mu=\text{max}(\mu_0,\mu_T)$. In both regimes, the magnetoresistance has a quadratic dependence on the magnetic field. Thus, particle-particle pairs with singlet symmetry contribute to the magnetoresistance as the conduction electrons would do, typical of the conventional Landau Fermi liquid theory~\cite{AlloulPRB11}.

\subsubsection{Spin-one case}
Next, we consider the situation where the spins of the particle-particle pairs have a triplet symmetry. In this scenario,
the boson scattering off conduction electrons generates a mass-correction due to the Zeeman field $H$ given by
\begin{align}
\mu =  \mu_0 +\mu_T +\mu_H,
\label{eq:muH_1}
\end{align}
where $\mu_H=\alpha H$ and $\alpha$ is a constant. For a comprehensive evaluation of
this mass renormalization, refer to Appendix~\ref{App:Triplet}. Again, the regimes will be determined by the maximum among $\mu_0$, $\mu_T$, and $\mu_H$. As a result, we have a regime where the mass-term couples linearly to the magnetic field. 
Taking the limit ${\gamma T/\mu \ll 1}$ in Eq.~(\ref{sigma2}), it is clear that the orbital contribution becomes negligible in this regime. Hence, the spin-one contribution to the magnetoresistance becomes 
\begin{align}
\dfrac{\Delta \rho_{xx}}{\rho_{xx}(0)} =  \begin{cases}
\dfrac{\alpha}{\mu_0+\mu_T} H & \mbox{when   } \text{max}(\mu_0,\mu_T,\mu_H)=\mu_H ,\\ \\
 \dfrac{\kappa}{\beta} H^2 & \mbox{otherwise   }  ,
\end{cases} 
\label{eq:MR_app_1}
\end{align}
where $\kappa/\beta\equiv-32 \gamma^2 T^2/( 5 \mu^4)$, with $\mu=\text{max}(\mu_0,\mu_T)$. For a detailed evaluation of all these quantities, refer to Appendix~\ref{App:MR_Triplet}. 

Note that $\mu_H \gg \mu_T$ can be recast in the form $H \gg \eta T$. Here, $\eta$ can be considered as a constant prefactor up to a logarithmic corrections\footnote{$\eta=\frac{\mu_0+\tilde{g}_b \log(\gamma T/\mu_0)}{\alpha}$}. Consequently, in this high field regime, we have a linear-in-$H$ magnetoresistance.   However, in the low field regime, $H\ll \eta T$, we have a quadratic $H$ dependence of the magnetoresistance. In a recent study~\cite{Hussey2020} on overdoped cuprates, the high field regime has a linear dependence of the MR with the magnetic field and displays a quadratic evolution of the MR at the low field regime. Remarkably, our calculations unveil that the incoherent bosons can explain such a behavior of the MR.

Lastly, we comment on the scaling of the in-plane magnetoresistance with that observed experimentally. The in-plane MR is given by
\begin{equation}
\Delta \rho_{xx} = \rho_{xx}(H,T)-\rho_{xx}(0,0).
\label{Quadmain1}
\end{equation}
Near the QCP, $\Delta \rho_{xx}$ follows a quadrature dependence~\cite{hayes2016scaling,Hayes18,Hussey2020,PhilipsMR19}, i.e.,
$\Delta \rho_{xx} = \sqrt{a^2 T^2 + b^2 H^2}$, where $a$ and $b$ are constants. In the low-field and high-field limits, this quantity can be easily obtained. 
We calculate the same for our theory, while restricting our attention to the case when the bosons emerge from pairs of electrons that have spin-triplet symmetry. Consequently, the mass-term is given by Eq.~(\ref{eq:muH_1}).  Again, as we mentioned before, the maximum among $\mu_0$, $\mu_T$, and $\mu_H$ determines the different regimes in the present theory. Therefore, the leading order contribution to this quantity becomes
\begin{align}
\Delta \rho_{xx} \propto  \begin{cases}
H & \mbox{for   } H \gg \eta T ,\\
\dfrac{H^2}{T} & \mbox{for }   H \ll \eta T,
\end{cases} 
\label{eq:MR_app_main121}
\end{align}
where, up to logarithmic corrections, $\eta$ is just a constant. The detailed evaluation is
presented in Appendix~\ref{App:ChangeinMR}. As a result, although our calculations cannot determine exactly the quadrature
dependence for $\Delta \rho_{xx}$, a similar scaling behavior is found in the low-field and high-field limits.

\section{Discussion\label{subsec:Discussion}}
This paper provides an intuitive model that accounts for the
multiple transport anomalies in the strange metal phase. The universal observations of charged or
neutral modulated excitations in the cuprates~\cite{Hamidian16, Hoffman02} along with the closeness of
finite momentum Cooper pairing to the uniform pairing state~\cite{Dai18,agterberg2020physics,tJPDW0,Freire15,Carvalho16} strongly hints at
remnant PDW fluctuations in the SM phase.
Initially, it is suggested that the pseudogap is
a transition towards a ``fluctuating'' pair density wave (PDW) phase~\cite{agterberg2020physics, Dai19},
which could readily lead to the
presence of charge-two finite momentum bosons in the strange metal phase.
Another recent proposal suggests that the pseudogap can result from fractionalizing 
a PDW state~\cite{Grandadam19, Chakraborty19}. Here, the gap opening at $T^{*}$ results
from a de-confining transition of a PDW order parameter into an SC
and charge density wave fields. The fluctuations of the gauge field associated with
the fractionalization produce the pseudogap. At $T=0$, this involves
a coherent superposition of particle-particle and particle-hole orders.
Here again, preformed PDW pairs can exist above $T^{*}$.
Several microscopic models~\cite{agterberg2020physics,tJPDW0,berg2010pair,peng2020evidence} are also
introduced to examine the possibility of such PDW sates.
In the presence of either time-reversal or parity symmetry, the strong correlation between
electrons becomes an essential ingredient for the generation of the PDW states~\cite{agterberg2020physics,yang2009nature}.
Nevertheless, these PDW pairs are typically expected to have a singlet spin symmetry. A few 
recent studies explore the feasibility of the PDW states
in the triplet channel as well~\cite{KimSci_19,tripletPDW20,Miranda21}, and
some proposals have suggested to fractionalize a stripe~\cite{nussinov2002_1,zaanen2003} or 
a spin density wave order~\cite{Sachdev_SDW_fraction}.

Consequently, on top of usual fermionic carriers,
we invoked the presence of charge-two bosonic excitations emerging from such fluctuating finite momentum Cooper pairs
in the SM phase. We show that such bosons contribute to the linear-in-$T$ resistivity and lead to a
broad Drude component in the optical conductivity with the dissipation of ``Planckian''-type. 
Since the bosons are incoherent, they do not
contribute to the Hall conductivity, thereby explaining the missing number
of carriers reported in the cuprates in the regions where longitudinal resistivity
is linear-in-$T$~\cite{putzke2019reduced,PRBGreven20}.
If bosons emerge from spin-one pairs of fermions, they also produce a linear-in-$H$ magnetoresistance.
Of course, our model also contains fermions, which provide
an additional part of the transport. The scattering around the Fermi
surface has to show a form of anisotropy in the transport lifetime,
according to which both the c-axis magnetoresistance~\cite{RossMcKenziePRL11,RossMcKenzie12,hussey2013generic,hussey2003coherent} and the fermionic quasiparticle
lifetime (extracted from the cotangent of the Hall angle~\cite{clayhold1989hall,Hall_Manko_92} $\cot\theta_{H}\sim\tau_{H}^{-1}\sim T^{2}$)
can be successfully reproduced. 

As a final remark, we note that since incoherent charge-two bosons contribute
to the Drude peak observed in the optical conductivity, these pairs could
also be a good candidate for explaining the missing spectral weight in the superfluid
density that is present in this region
of the phase diagram~\cite{lobo2011optical,HomesSumRule,LeeSinova08}.

\section*{Acknowledgement}
The authors thank Saheli Sarkar, Nigel Hussey, Yvan Sidis, Dmitrii Maslov, and Debmalya Chakraborty for valuable discussions. This work has received financial support from the ERC, under grant agreement AdG694651-CHAMPAGNE. H.F. acknowledges funding from CNPq under Grants No. 405584/2016-4 and No. 310710/2018-9.

\appendix
\section{Scattering through fermions}
\label{App:Scattering}
\begin{figure}[b]
\includegraphics[width=0.45\textwidth]{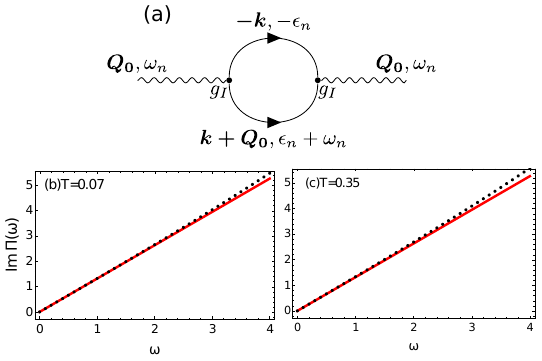}
\caption[0.5\textwidth]{\textbf{(a)} The leading order boson propagator correction, given by Eq.~\eqref{eq:SM1}. The solid line is the bare electronic Green's function, $\mathcal{G}$.  The wavy lines are the finite momentum bosons with ordering wave-vector $\mathbf{Q_0}$. \textbf{(b)} Comparison of imaginary part of $\Pi(\omega)$ for numerical and approximate analytical evaluations for low temperature, $T=0.07$ and $g_I=1$, $Q_0=\pi/2$ \text{(b)} Same for the higher temperature $T=0.35$. The following physical constants are set to unity: $\hbar=1$, $k_B=1$, and $e=1$.}
\label{scatter}
\end{figure}

In this section, we formally show that the scattering through fermions leads to a diffusive imaginary part of the self-energy of finite momentum bosons. Fig.~(\ref{scatter}a) shows the relevant Feynman diagram. The bosons emerge from the pairs of fermions with finite-momentum $\mathbf{Q_0}$. The wavy-lines represent the bosons, and the solid lines denote the fermions.
The expression for the diagram reads as
\begin{eqnarray}
\Pi(\omega_{n},\mathbf{Q}_0)  = &\dfrac{g^2_I}{L}  \displaystyle \sum_{\mathbf{k}}  T \displaystyle \sum_{\varepsilon_n} \mathcal{G}(-\varepsilon_n,-\mathbf{k}) \mathcal{G}(\varepsilon_n+\omega_n,\mathbf{k}+ \mathbf{Q_0}) \nonumber \\ &+\mathcal{G}(-\varepsilon_n,-\mathbf{k}) \mathcal{G}(\varepsilon_n-\omega_n,\mathbf{k}-\mathbf{Q_0}).
\label{eq:SM1}
\end{eqnarray}
Here $L$ is the volume of of the system, $T$ is the temperature and $g_I$ is the interaction strength between the finite momentum bosons and fermions. The frequencies, $\varepsilon_n$ and $\omega_n$ are fermionic and bosonic Matsubara frequencies, respectively. The Green's functions, $\mathcal{G}$, denote the free fermionic propagators given by
\begin{eqnarray}
\mathcal{G}^{-1}({\mathbf{k},\omega_{n}}) & = i \epsilon_n -\xi_\mathbf{k},
\label{eq:SM2}
\end{eqnarray}
where $\xi_\mathbf{k}=\hbar^2 k^2/2 m_e$. For simplicity of notations, we set $\hbar^2 /2 m_e=1$, from now on. In order to perform the Matsubara summation, we go to the complex plane by performing the substitution, $i \epsilon_n \rightarrow z$. The first term of the RHS of Eq.~(\ref{eq:SM1}) becomes
\begin{eqnarray}
\Pi(\omega_{n},\mathbf{Q}_0) & = - \dfrac{g^2_I}{L} \displaystyle \sum_{\mathbf{k}} \displaystyle \oint_{\mathcal{C}} \dfrac{dz}{2 \pi i} \dfrac{n_F(z)}{(z+\xi_{-\mathbf{k}})(z+i \omega_n -\xi_{\mathbf{k}+\mathbf{Q_0}})}.\nonumber\\
\label{eq:SM3}
\end{eqnarray}
The integral  is evaluated using the residue theorem and obtain
\begin{eqnarray}
\Pi(\omega_{n},\mathbf{Q}_0) & = - g^2_I \dfrac{1}{L}  \displaystyle \sum_{\mathbf{k}}  \dfrac{1-n_F(\xi_{-\mathbf{k}})-n_F(\xi_{\mathbf{k}+\mathbf{Q_0}})}{i \omega_n-\xi_{-\mathbf{k}} -\xi_{\mathbf{k}+\mathbf{Q_0}}}.\nonumber\\
\label{eq:SM4}
\end{eqnarray}
We perform the analytic continuation by letting ${i \omega_n \rightarrow \omega + i 0^+}$ and then taking the imaginary part
\begin{eqnarray}
\text{Im } \Pi(\omega,\mathbf{Q}_0) = &\frac{ \pi g^2_I}{L} \displaystyle \sum_{\mathbf{k}} \left[1-n_F(\xi_{-\mathbf{k}})-n_F(\xi_{\mathbf{k}+\mathbf{Q_0}}) \right] \nonumber \\
 &\times \delta\left(\omega-\xi_{-\mathbf{k}} -\xi_{\mathbf{k}+\mathbf{Q_0}}\right).
\label{eq:SM5}
\end{eqnarray}
The $\mathbf{k}$-summation is performed by converting it to an integral. We can approximate $\xi_{\mathbf{k}+\mathbf{Q_0}} \approx k^2+Q_0^2+2 k_F Q_0 \cos(\theta)$, where $\theta$ is the angle between Fermi-momentum $\mathbf{k_F}$ and the ordering wave-vector, $\mathbf{Q_0}$. Furthermore, we use the flat-band approximation with the density of states at the Fermi energy given by $\mathcal{N}(\epsilon_F)$, the integral in two dimensions becomes
\begin{widetext}
\begin{align}
\text{Im } \Pi(\omega,\mathbf{Q}_0) = \dfrac{ g^2_I \mathcal{N}(\epsilon_F)}{16 \pi} \displaystyle \int^{2 \pi}_{0} d\theta 
\left[\tanh\left(\frac{\omega+Q_0^2+2k_F Q_0\cos(\theta)}{4T}\right) +\tanh\left(\frac{\omega-Q_0^2-2k_F Q_0\cos(\theta)}{4T}\right) \right].
\label{eq:SM7}
\end{align}
In the limit, $T\rightarrow 0$, we can approximate $\tanh(x/T)\approx \text{sgn}(x)$. In this low-temperature regime, the integrand in the square brackets in Eq.~(\ref{eq:SM7}), which we simply denote as  $I(\theta)$ from now on, is approximately given by
\begin{align}
I(\theta) = \begin{cases} 
2 & \mbox{if } \theta \in \left[ \cos^{-1}\left(\frac{\omega-Q_0^2}{2 k_F Q_0}\right) , \cos^{-1}\left(\frac{-\omega-Q_0^2}{2 k_F Q_0}\right) \right],\\ \\
2 & \mbox{if } \theta \in \left[ 2 \pi- \cos^{-1}\left(\frac{-\omega-Q_0^2}{2 k_F Q_0}\right), 2 \pi - \cos^{-1}\left(\frac{\omega-Q_0^2}{2 k_F Q_0} \right) \right], \\ \\
0 & \mbox{otherwise }.\end{cases} 
\end{align}
The form of $I(\theta)$ is used to evaluate the integral in Eq.~(\ref{eq:SM7}) and it reads as
\begin{eqnarray}
\text{Im } \Pi(\omega,\mathbf{Q}_0)  =  \frac{ g^2_I \mathcal{N}(\epsilon_F)}{4 \pi} \left[ \cos^{-1}\left( \frac{-\omega-Q_0^2}{2 k_F Q_0} \right)- \cos^{-1}\left( \frac{\omega-Q_0^2}{2 k_F Q_0} \right) \right].
\label{eq:SM8}
\end{eqnarray}
\end{widetext}
Finally, expanding the function for $\omega \ll 2 k_F Q_0$, we arrive at the result
\begin{eqnarray}
\text{Im } \Pi(\omega,\mathbf{Q}_0) & = & \frac{ g^2_I \mathcal{N}(\epsilon_F)}{2 \pi} \frac{\omega}{\sqrt{(2 k_F Q_0)^2-Q_0^4}}.
\label{eq:SM9}
\end{eqnarray}
This shows there is a linear dependence on $\omega$. Performing similar calculations for the second term in Eq.~(\ref{eq:SM1}) and the imaginary part of the self-energy reads
\begin{eqnarray}
\text{Im } \Pi(\omega,\mathbf{Q}_0) & = & \gamma \vert \omega \vert,
\label{eq:SM10}
\end{eqnarray}
with $\gamma=\frac{ g^2_I \mathcal{N}(\epsilon_F)}{2 \pi\sqrt{(2 k_F Q_0)^2-Q_0^4}}$. We have checked our approximate expression against numerical evaluation of Eq.~(\ref{eq:SM7}). A good agreement between them is observed in Fig.~(\ref{scatter}b) at low temperature, and in Fig.~(\ref{scatter}c) at high temperature.

\subsection{Landau damping for electrons near the hotspots}
\label{App:LDHS}
Here we show that the particle-particle bubble evaluated in the previous section gives a Landau damped form if the electrons lives near the hotspots, as shown in Fig.~(1 a) of the main text. In the top two hot spots of the same figure the Fermi velocity along the $x$ and $y$ direction changes from $(-v_x,v_y) \rightarrow (v_x,v_y)$. Consequently, the dispersion becomes
\begin{align}
&\xi_{\mathbf{l}} = -l_x v_x+l_y v_y,\\
&\xi_{\mathbf{l}+\mathbf{Q}_0} = l_x v_x + l_y v_y.
\end{align}
Putting these two in Eq.(\ref{eq:SM5}) we obtain
\begin{align}
\text{Im } \Pi(\omega,\mathbf{Q}_0) = &\frac{ \pi g^2_I \text{sgn}(\omega)}{4 \pi^2} \int_{-\infty}^{\infty} dl_x \int_{-\infty}^{\infty} dl_y \delta\left(\omega-2 l_x v_x\right) \nonumber \\ &\times \left[1-n_F(l_x v_x-l_y v_y)-n_F(l_x v_x + l_y v_y) \right].
\label{eq:HS1}
\end{align}
Defining $\tilde{l}=v l$, after simplification we obtain
\begin{align}
%\text{Im } \Pi(\omega,\mathbf{Q}_0) &= \frac{ g^2_I \text{sgn}(\omega)}{8 \pi v_x v_y} \int_{-\infty}^{\infty} d\tilde{l}_x \int_{-\infty}^{\infty} d\tilde{l}_y  \left[\tanh\left(\frac{\tilde{l}_x-\tilde{l}_y}{2T}\right)+\tanh\left(\frac{\tilde{l}_x+\tilde{l}_y}{2T}\right) \right] \delta\left(\omega-2 \tilde{l}_x\right),\\
\text{Im } \Pi(\omega,\mathbf{Q}_0) = \frac{ g^2_I \text{sgn}(\omega)}{16 \pi v_x v_y} &\int_{-\infty}^{\infty} d\tilde{l}_y  \left[\tanh\left(\frac{\omega/2-\tilde{l}_y}{2T}\right) \right. \nonumber \\
 & \left.+ \tanh\left(\frac{\omega/2+\tilde{l}_y}{2T}\right) \right].
\label{eq:HS3}
\end{align}
In the limit, $T\rightarrow 0$, we can approximate $\tanh(x/T)\approx \text{sgn}(x)$. In this low-temperature regime, the integrand in the square brackets in Eq.~(\ref{eq:HS3}), has a constant value of $2$, when $\tilde{l}_y$ is restricted between $(-\omega/2,\omega/2)$, otherwise it vanishes. Therefore, performing the integration over $\tilde{l}_y$, we get
\begin{align}
\text{Im } \Pi(\omega,\mathbf{Q}_0) &=\frac{ g^2_I}{8 \pi v_x v_y} \vert \omega \vert.
\label{eq:HS4}
\end{align}
Consequently, we obtained Landau damping for the electrons near the hot-spots. 

\subsection{Robustness of Landau damping}
\label{App:RobLD}
We use a generalized form of the fermionic self-energy and show that the previously obtained Landau damping form is robust against such perturbations. Suppose this fermionic self-energy arises from a different physical mechanism, which is not considered in this paper. Following the notations in Ref.~\cite{Metlitski10b}, we assume the self energy of form $\Sigma_{l_{\tau}}=\vert \Sigma(l_\tau) \vert \text{sgn}(l_\tau)$. Next, we estimate the particle-particle bubble
\begin{align}
\Pi(q_\tau,\mathbf{Q}_0) =\frac{ i g^2_I }{8 \pi^3 v_x v_y} \int_{-\infty}^{\infty} dl_\tau\int_{-\infty}^{\infty} d\tilde{l}_x \int_{-\infty}^{\infty} d \tilde{l}_y \nonumber \\ 
 \times \frac{1}{\left( i \Sigma_{l_\tau}+\tilde{l}_x-\tilde{l}_y\right)\left( i \Sigma_{l_\tau+q_\tau}-\tilde{l}_x-\tilde{l}_y\right)}.
\label{eq:Rbs1}
\end{align}

If $q_{\tau}>0$ the poles of $\tilde{l}_y$ are in the opposite half-planes if the $l_\tau$ is restricted between $-q_\tau \leq l_\tau \leq 0$. We close the contour in the upper half-plane and obtain
\begin{align}
\Pi(q_\tau>0,\mathbf{Q}_0) =&-\frac{  g^2_I }{4 \pi^2 v_x v_y} \int_{-q_\tau}^{0} dl_\tau\int_{-\infty}^{\infty} d\tilde{l}_x  \nonumber \\ &\times \frac{1}{\left( i \Sigma_{l_\tau}-i \Sigma_{l_\tau+q_\tau}+2 \tilde{l}_x\right)},\\
\Pi(q_\tau>0,\mathbf{Q}_0) =&-\frac{ g^2_I }{8 \pi^2 v_x v_y} \int_{-q_\tau}^{0}  dl_\tau \nonumber \\ &\times  \log\left(\frac{i \Sigma_{l_\tau}-i \Sigma_{l_\tau+q_\tau}+2 \Lambda}{i \Sigma_{l_\tau}-i \Sigma_{l_\tau+q_\tau}-2 \Lambda }\right),
\label{eq:Rbs2}
\end{align}
where $\Lambda$ is the UV cutoff. If $\Sigma_{l_\tau}-\Sigma_{l_\tau+q_\tau} \ll 2\Lambda$, then logarithm can be approximated as $-i\pi$. The imaginary part of the $\Pi$ then becomes
\begin{align}
\text{Im } \Pi(q_\tau>0,\mathbf{Q}_0) &=\frac{ g^2_I }{8 \pi v_x v_y} q_\tau.
\label{eq:Rbs3}
\end{align}
Similarly one can repeat the procedure for $q_\tau<0$ and obtains the same expression with a negative sign. Therefore, combining these two one can write
\begin{align}
\text{Im } \Pi(q_\tau,\mathbf{Q}_0) &=\frac{ g^2_I }{8 \pi v_x v_y} \vert q_\tau \vert.
\label{eq:Rbs4}
\end{align}
Therefore, the Landau damping remains unaffected for arbitrary self-energy corrections. These conclusions remain unaffected if the electrons attain mass away from the putative hot-spots. To recognize this, we replace, in Eq.~(\ref{eq:Rbs1}), with $\mu_1$ and $\mu_2$ as the general mass of the electrons,
\begin{align}
\Pi(q_\tau,\mathbf{Q}_0) =&\frac{ i g^2_I }{8 \pi^3 v_x v_y} \int_{-\infty}^{\infty} dl_\tau\int_{-\infty}^{\infty} d\tilde{l}_x \int_{-\infty}^{\infty} d \tilde{l}_y  \nonumber \\
 &\times \frac{1}{\left( i \Sigma_{l_\tau}-\mu_1+\tilde{l}_x-\tilde{l}_y\right)\left( i \Sigma_{l_\tau+q_\tau}+\mu_2-\tilde{l}_x-\tilde{l}_y\right)}.
\label{eq:Rbs5}
\end{align}
Using the same procedure as above we arrive at
\begin{align}
\Pi(q_\tau>0,\mathbf{Q}_0) =&-\frac{ g^2_I }{8 \pi^2 v_x v_y} \int_{-q_\tau}^{0}  dl_\tau \nonumber \\ &\times \log\left(\frac{i \Sigma_{l_\tau}-i \Sigma_{l_\tau+q_\tau} -\mu_1 - \mu_2+2 \Lambda}{i \Sigma_{l_\tau}-i \Sigma_{l_\tau+q_\tau}-\mu_1 - \mu_2 -2 \Lambda }\right).
\label{eq:Rbs6}
\end{align}
Again, if $\Sigma_{l_\tau}-\Sigma_{l_\tau+q_\tau} \ll 2\Lambda-\mu_1-\mu_2$, we have the same Landau damping form as found in Eq.~(\ref{eq:HS4}).

\subsection{On the fermion-boson vertex corrections}
\label{App:FBVertex}
Recent studies of the antiferromagnetic QCP~\cite{Metlitski10b,lee18NFL} in two spatial dimensions obtained that the fermion-boson vertex corrections become relevant at low-energy scales and modify the dynamical exponent close to the QCP. Therefore, it becomes essential to discuss these vertex corrections in the present case. Our calculation for the finite-momentum bosons will follow the results of Ref.~\cite{Metlitski10b}, and we present it here for completeness. In the present study, two different situations can emerge. 

\begin{figure}
\centering
      \includegraphics[width=8.5cm,keepaspectratio]{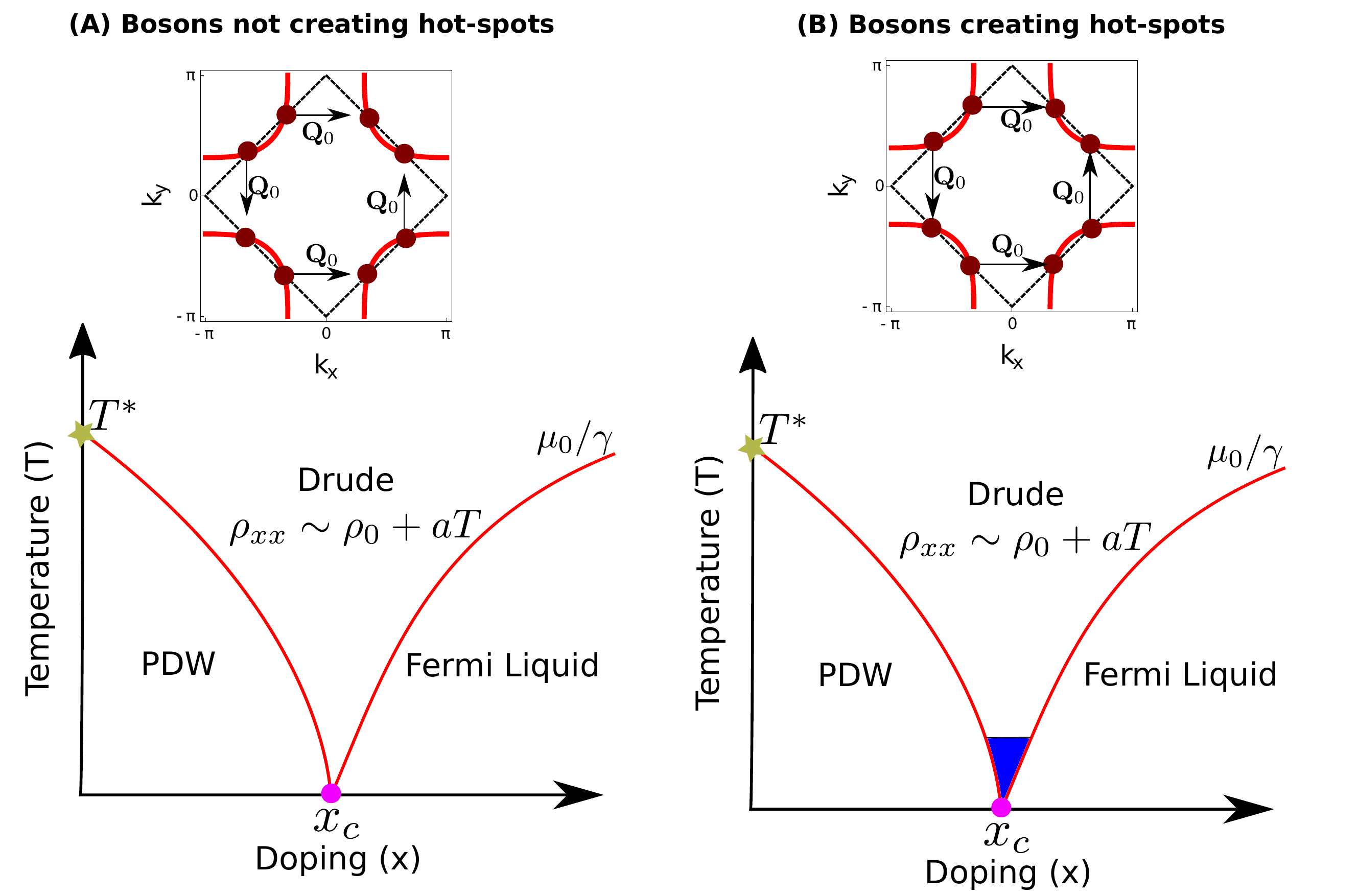}
      \caption{Panel (A) on the left side depicts the scenario when the bosons do not create hot-spots since the bosonic wave-vector $\mathbf{Q}_0$ is smaller than the distinct parts of the Fermi surface. In this scenario, the linear-in-$T$ behavior of resistivity and Drude form of optical conductivity extends to $T=0$ at the critical dopings.  Panel (B) on the right side presents the scenario when the bosons create hot-spots by connecting the Fermi-surface. In this scenario, the fermion-boson vertex correction becomes relevant and changes the dynamical exponent of the QCP. However, this happens only at low temperatures near the QCP here represented by the blue region. Above this temperature, we can have the same linear-in-$T$ resistivity with Drude conductivity for a broad temperature regime.}  
      \label{fig1_1}
\end{figure}

In the first one, the bosons cannot generate hot-spots if the wave-vector $\mathbf{Q}_0$ is either too small or too large to connect distinct parts of the Fermi surface as shown in Fig.~(\ref{fig1_1}A). Thus, the fermionic propagator reestablishes the Fermi liquid behavior of Eq.~(\ref{eq:SM2}). In this scenario, the Landau damping remains unchanged, whereas the vertex corrections become irrelevant. Hence, the transport properties of the model give the $T$-linear behavior of the resistivity and a broad Drude component extending to zero temperature as exhibited in the phase-diagram of Fig.~(\ref{fig1_1}A).

In the second situation, if the bosonic wave-vector $\mathbf{Q}_0$ create hot-spots by connecting two distinct parts of the Fermi surface as displayed in Fig.~(\ref{fig1_1}B), the fermionic self-energy is given by
\begin{align}
\text{Im }\Sigma(\omega,\mathbf{Q}_0) = C \vert \omega \vert^{1/2} \text{sgn}(\omega),
\label{eq:VCe1}
\end{align}
where $C$ is just a constant, and the self-energy has a Non-Fermi liquid behavior. Additionally, in this situation, the vertex corrections also become relevant. The integral to evaluate the same is given by
\begin{align}
\Gamma(0,0) =&\frac{ i }{8 \pi^3 v_x v_y} \int_{-\infty}^{\infty} dl_\tau\int_{-\infty}^{\infty} d\tilde{l}_x \int_{-\infty}^{\infty} d \tilde{l}_y \nonumber \\ &\times  \frac{1}{\left( i \Sigma_{l_\tau}+\tilde{l}_x-\tilde{l}_y\right)\left( i \Sigma_{l_\tau}-\tilde{l}_x-\tilde{l}_y\right)(\gamma \vert l_\tau \vert +\tilde{l}_x^2+ \tilde{l}_y^2)}.
\label{eq:VCe2}
\end{align}
After computing this integral, one obtains a logarithmic divergence from the vertex corrections, thereby affecting the dynamical exponent near the QCP. However, recent sign-problem-free quantum Monte Carlo studies suggest that only within a small temperature regime near the QCP~\cite{gerlach17} (which is, in fact, too low to be seen in these simulations), these vertex corrections would become relevant which is represented by the blue region of phase-diagram in Fig.~(\ref{fig1_1}B). Furthermore, before reaching such a low-temperature regime, we point out that this divergence can also be regularized by other mechanisms of damping, such that $\Sigma_{\omega} \gg  C \vert \omega \vert^{1/2} \text{sgn}(\omega)$. In the cuprates these other sources of damping can have many different origins, such as nematic fluctuations, loop-current fluctuations~\cite{Xavier15}, among others. These additional fluctuations that emerge in these materials can also regularize the fermion-boson vertex without changing the transport properties.

\section{Renormalization of the ``mass'' term -- Number of bosons}
\label{App:Nb}
\begin{figure}[h!]
\includegraphics[width=0.45\textwidth]{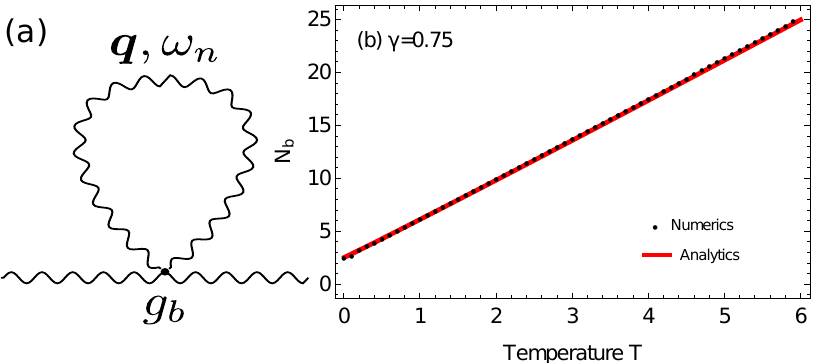}
\caption[0.5\textwidth]{\textbf{(a)} The first-order diagram of the bosonic self-energy. The wavy lines denote the bosons, which interact with the strength $g_b$. \textbf{(b)} The temperature dependence of the number of bosons, evaluated by solving the Eq.~(\ref{EqNbnum}) numerically and compared with the expression arrived at analytically in Eq.~(\ref{eqNb1}). The perfect match between the two evaluations gives us confidence in our analytical results. The temperature is in the units of $\mu_0/\gamma$. The following physical constants are set to unity: $\hbar=1$, $k_B=1$, and $e=1$.}
\label{figNb}
\end{figure}
In this section, we present the detailed evaluation of the leading order term in the self-energy, which renormalizes the mass of the bosons. Fig.~(\ref{figNb}a) shows the relevant diagram, where the wavy-lines represent the bosons, which interact with other bosons with the interaction strength being represented by $g_b$. The mass term renormalization is given by the real part of this diagram, i.e.
\begin{equation}
N_b=\frac{1}{L} \sum_{\mathbf{q}} T \sum_{\nu_n} \frac{1}{\gamma \vert \nu_n \vert + q^2+ \mu}.
\end{equation}
The Matsubara summation over $\varepsilon_n$ is carried out by using the spectral decomposition of the bosonic Green's function. The spectral function $\mathcal{A}(E,\mathbf{q})$ is given by~\cite{altlandBook}
\begin{equation}
\mathcal{A}(\mathbf{q},E)=-2 \text{Im }[\mathcal{D}_R(\mathbf{q},E)]=-2\frac{\gamma E}{(\gamma E)^2+(q^2+\mu)^2}.
\label{Spectral1}
\end{equation}
Noting that $\mathcal{D}(\mathbf{q},\nu_n)=\int^{\infty}_{-\infty}\frac{dE}{2 \pi}\frac{\mathcal{A}(\mathbf{q},E)}{i\nu_n-E}$, the summation is taken to the complex plane by promoting $i \nu_n \rightarrow z$ and $T\sum{\nu_n} \rightarrow \oint_{\mathcal{C}} \frac{dz}{2 \pi i} n_B(z)$, where $\mathcal{C}$ covers the whole of the complex plane. Therefore the expression becomes
\begin{align}
\nonumber N_b &= \frac{1}{L} \sum_{\mathbf{q}} \oint_{\mathcal{C}} \frac{dz}{2 \pi i} \int^{\infty}_{-\infty}\frac{dE}{2 \pi}\frac{\mathcal{A}(\mathbf{q},E) n_B(z)}{z-E},\\
N_b&=-\frac{1}{2\pi} \int^{\infty}_{0}  dq \int^{\infty}_{-\infty}\frac{dE}{2 \pi} n_B(E) \frac{\gamma E}{(\gamma E)^2+(q^2+\mu)^2}.
\end{align}
After performing the integral over $\mathbf{q}$ exactly, $N_b$ becomes
\begin{align}
N_b&=-\frac{1}{4\pi^2} \int^{\infty}_{-\infty} dE  \left[ \frac{\pi}{2}\text{sgn}(E)-\tan^{-1}\left(\frac{\mu}{\gamma T}\right)\right] n_B(E).
\label{EqNbnum}
\end{align}

The Bose-Einstein distribution is approximated as
\begin{align}
n_B(x) = \begin{cases} 
0 & \mbox{if } x>T, \\
\dfrac{T}{x} & \mbox{if } \vert x \vert < T, \\ \\
- 1 & \mbox{if } x<-T. 
\end{cases} 
\label{apprxnb}
\end{align}
Performing the integral in the regime where $\vert E \vert <T$, the renormalization of the mass term reads
\begin{align}
N^{(1)}_b = \begin{cases} 
\dfrac{T}{4 \pi }\log\left(\frac{\gamma T}{\mu}\right) & \mbox{for   } \gamma T \gg \mu, \\ \\
\dfrac{\mu}{2 \pi^2 \gamma} & \mbox{for   } \gamma T \ll \mu.
\end{cases} 
\end{align}
Similarly, performing the integral for $E<-T$, we obtain
\begin{align}
N^{(2)}_b = \begin{cases} 
\dfrac{1}{4 \pi } (\Lambda-T) & \mbox{for   } \gamma T \gg \mu, \\ \\
0 & \mbox{for   } \gamma T \ll \mu,
\end{cases} 
\end{align}
where $\Lambda$ is the ultraviolet energy cutoff of the system. Therefore, $N_b$ will be independent of temperature in this regime, as $\Lambda$ will be the dominant energy scale. This gives the number of bosons that condenses to the ground state. The mass term $\mu$ to the first order is given by setting $\mu=\mu_0$, where $\mu_0$ is the bare mass of the bosons, which is naturally temperature independent.  Therefore, to first order in $g_b$, we obtain
\begin{align}
\mu = \begin{cases} 
\mu_0+g_b \left( \dfrac{T}{4\pi}\log\left( \frac{\gamma T}{\mu_0}\right) \right) & \mbox{for   } \gamma T \gg \mu_0, \\ \\
\mu_0 & \mbox{for   } \gamma T \ll \mu_0.
\end{cases} 
\label{eqNb1}
\end{align}
The constant terms are absorbed in the $\mu_0$, which becomes close to zero near the quantum critical point. The temperature dependence of $N_b$ calculated numerically from Eq.~(\ref{EqNbnum}) and analytical form displayed in Eq.~(\ref{eqNb1}) matches over a wide range of temperature, as can be seen in Fig.~(\ref{figNb}b)

\section{Longitudinal conductivity: Kubo formula}
\label{App:Kubo}

\begin{figure}[h!]
\includegraphics[width=0.45\textwidth]{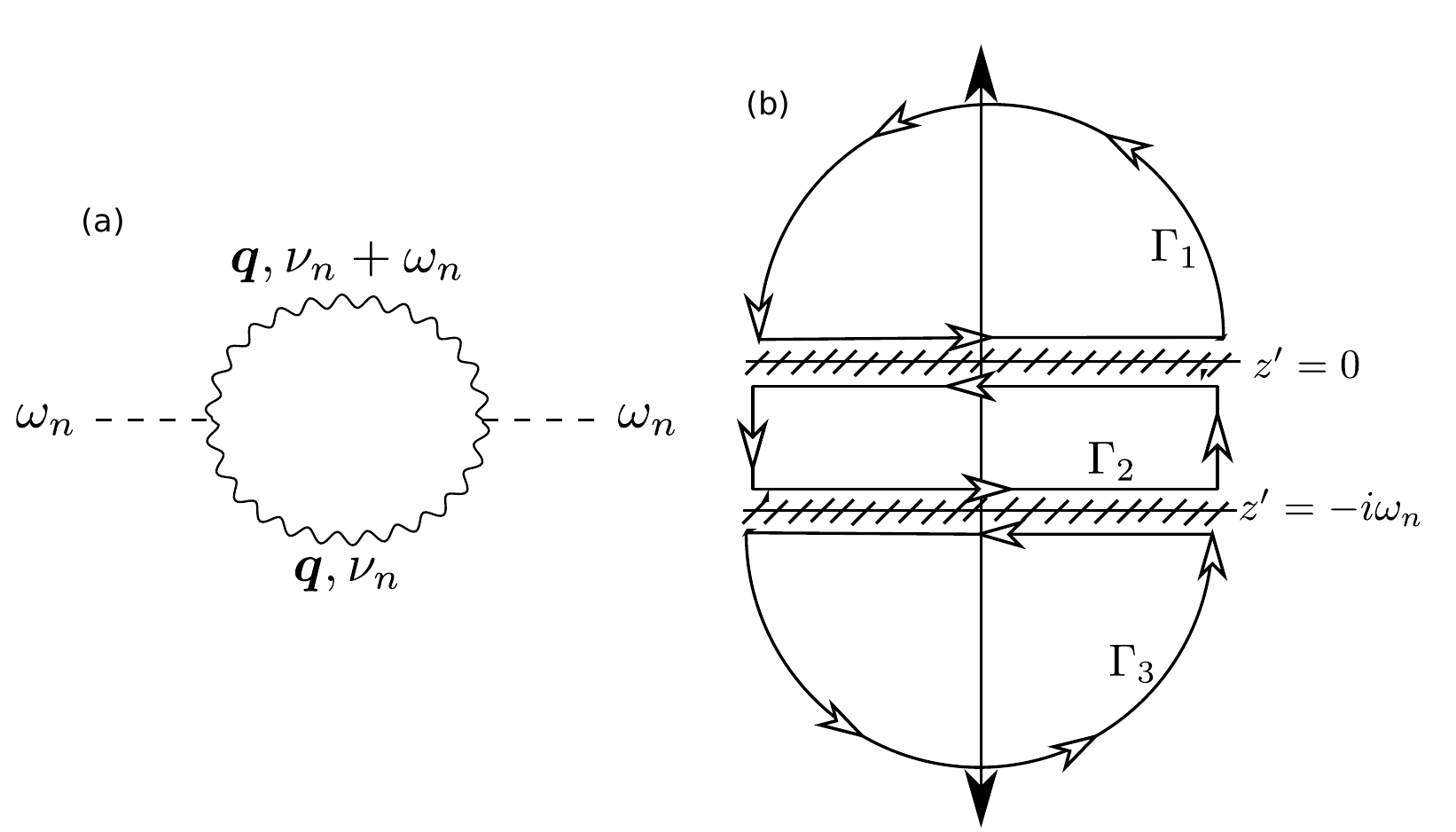}
\caption[0.5\textwidth]{\textbf{(a)} The leading order diagram to evaluate the conductivity. \textbf{(b)} The contours used to evaluate the Kubo formula for finite-momentum bosons. The two dashed lines are the branch cuts.}
\label{figSig}
\end{figure}

The longitudinal conductivity is given in terms of correlation functions $\mathcal{K}$ by~\cite{altlandBook}
\begin{align}
\mathcal{K}(\omega_n)=-T\sum_{\nu_n} \frac{1}{L} &\sum_{\mathbf{q}} \left[ \mathcal{D}(\nu_n,\mathbf{q})  + q^2 \mathcal{D}(\nu_n,\mathbf{q}) \mathcal{D}(\nu_n+\omega_n,\mathbf{q}) \right].
\label{Kxx}
\end{align}
The first term is the diamagnetic term and the second term is the paramagnetic current-current correlation. The finite-momentum bosons is dominant around $\mathbf{q}=\mathbf{Q_0}$. The Eq.~(\ref{Kxx}) can be approximated by
\begin{align}
\mathcal{K}(\omega_n)\approx-T\sum_{\nu_n} \frac{1}{L} & \sum_{\mathbf{q}} \left[ \mathcal{D}(\nu_n,\mathbf{q}) + Q_0^2 \mathcal{D}(\nu_n,\mathbf{q}) \mathcal{D}(\nu_n+\omega_n,\mathbf{q}) \right].
\label{Kxx2}
\end{align}
The optical conductivity is then evaluated by
\begin{equation}
\sigma(\omega)=-\frac{K(\omega_n)}{\omega_n}\Bigg\vert_{i \omega_n \rightarrow \omega+i0^+}.
\label{sigmaw}
\end{equation}

The evaluation of the $K$ is carried out in the following way: The integral is evaluated in the contour shown in Fig.~(\ref{figSig}b). There are two branch cuts -- at $z^{\prime}=0$ and $z^{\prime}=i\omega_n$. The integrals over the $\Gamma_1$ and $\Gamma_3$ contours cancel the diamagnetic term. Therefore, only the $\Gamma_2-$contour contributes to the optical conductivity. The integral becomes
\begin{align}
K(\omega_n)=\frac{-Q_0^2}{2 \pi i L} \sum_{\mathbf{q}}\oint_{\Gamma_2}
\frac{dz \,n_B(z)}{\left( i \gamma z + q^2 + \mu \right)\left( (-i z+\omega_n)\gamma  + q^2 + \mu \right)}.
\end{align}
The poles of $z$ lie outside the $\Gamma_2$-contour and hence the full integrals collapse to the real line integrals along the branch cuts. The resulting expression becomes
\begin{align}
K(\omega)=\frac{Q_0^2}{L} &\sum_{\mathbf{q}} \frac{1}{2 \pi i} \int_{-\infty}^{\infty} 
dx \nonumber \\ &\times \frac{n_B(x-\omega/2)-n_B(x+\omega/2)}{\left( i \gamma x - i \gamma \frac{\omega}{2}+q^2 +\mu \right)\left( -i \gamma x - i \gamma \frac{\omega}{2}+q^2 +\mu \right)}.
\end{align}
The summation over $\mathbf{q}$ is converted to an integral and it is performed by usual means, i.e.
\begin{align}
K(\omega)=-\frac{Q_0^2 \omega}{16 \pi^2 \gamma} &\int_{-\infty}^{\infty} 
dx  \nonumber \\
 &\times \left(\frac{\partial n_B}{\partial x} \right) \frac{1}{x} \log \left( \frac{-i \gamma x - i \gamma \frac{\omega}{2} +\mu}{i \gamma x - i \gamma \frac{\omega}{2}+\mu} \right).
\label{eq_sigma_num}
\end{align}
From the approximate form of the $n_B$ given in Eq.~(\ref{apprxnb}), we obtain
\begin{align}
\frac{\partial n_B}{\partial x} = \begin{cases} 
0 & \mbox{if } \vert x \vert >T, \\ \\
-\dfrac{T}{x^2} & \mbox{if } \vert x \vert < T. \end{cases}
\label{eqn:nb2} 
\end{align} 
Using Eq.~(\ref{eqn:nb2}), the optical conductivity becomes
\begin{equation}
\sigma(\omega) = -\frac{i Q_0^2 T}{16 \pi^2 \gamma}\int_{-T}^{T} 
dx \frac{1}{x^3} \log \left( \frac{-x - \frac{\omega}{2}- \frac{i \mu}{\gamma}}{x - \frac{\omega}{2}- \frac{i \mu}{\gamma}} \right),
\end{equation}
where we defined $\tilde{\mu} = \frac{\omega}{2}+ i \frac{\mu}{\gamma}$. As a result, performing the integral, we obtain,
\begin{align}
\sigma(\omega) = -\frac{i Q_0^2 T}{16 \pi^2 \gamma}  &\left[ -\frac{2}{\tilde{\mu} T} +\frac{1}{2 \tilde{\mu}^2} \log\left( \frac{\tilde{\mu}+T}{\tilde{\mu}-T}\right)  \right. \nonumber \\  
&\left.   -\frac{1}{2 \tilde{\mu}^2} \log\left( \frac{\tilde{\mu}-T}{\tilde{\mu}+T}\right) + \frac{1}{2T^2} \log \left( \frac{\tilde{\mu}-T}{\tilde{\mu}+T} \right) \right. \nonumber \\
 &\left. -\frac{1}{2T^2} \log \left( \frac{\tilde{\mu}+T}{\tilde{\mu}-T} \right) \right].
\label{Kubo_primitive}
\end{align}

We expand the above expression in two regimes: For the first regime, $T\ll \sqrt{\omega^2/4+ \mu^2/\gamma^2}$ we find that the optical conductivity
displays the Drude form
\begin{equation}
\sigma(\omega)=\frac{Q_0^2}{4\pi^2 \mu \left( 1- i\frac{\gamma \omega}{2 \mu} \right)}.
\label{sigma_ana}
\end{equation}
We have compared the static conductivity given in the above equation against the numerical evaluation for the same using Eq.~(\ref{eq_sigma_num}). This comparison is displayed in the main text. A remarkable match between the two computations over a wide range of temperatures is observed. The Drude conductivity is naturally given by: $\sigma(\omega)=\sigma_0 \frac{\tau}{1-i \omega \tau}$. From that expression, one can easily read off the ${\sigma_0=\frac{Q_0^2}{2 \pi^2 \gamma}}$ while the scattering time of the bosons is given by $\tau=\frac{\gamma}{2\mu}$. \\

On the other hand, for the second regime ${T\gg \sqrt{\omega^2/4+ \mu^2/\gamma^2}}$ the optical conductivity does not exhibit the traditional Drude form
\begin{equation}
\sigma(\omega)=\frac{Q_0^2 \mu }{12\pi^2 \gamma^2 T^2} \left( 1- i\frac{\gamma \omega}{2 \mu} \right).
\label{sigma0_1}
\end{equation}
In the next section, we discuss the temperature dependence of the dc conductivity.

\subsection{Static Conductivity -- The regimes}
\label{App:Static}

\begin{figure}[h!]
\includegraphics[width=0.45\textwidth]{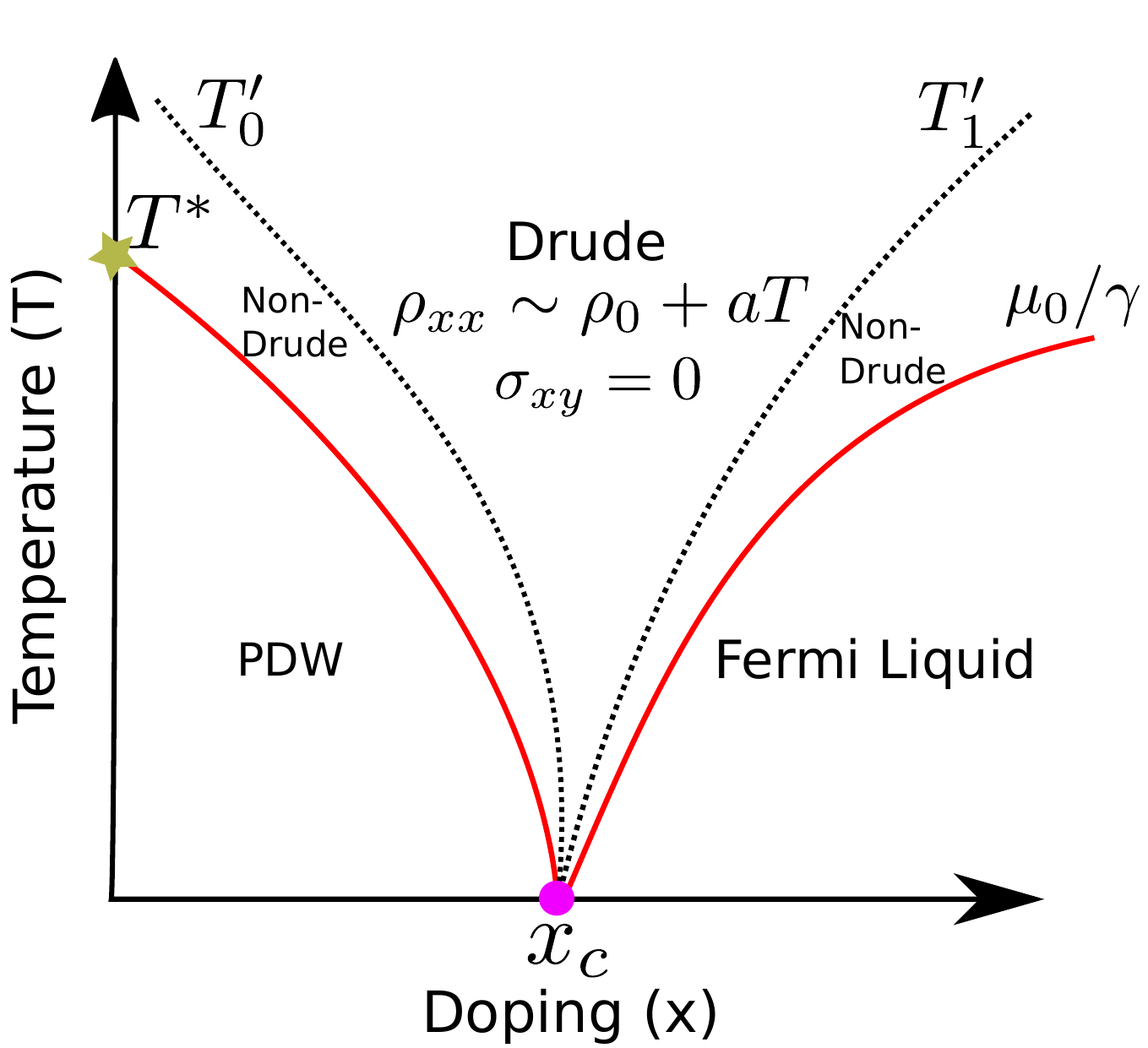}
\caption[0.5\textwidth]{The phase diagram for the scenario when bosonic interaction strength is weaker than the Landau damping parameter. Here, we have an intermediate regime bounded by the dotted black curve, where the optical conductivity does not conform to the conventional Drude form.}
\label{fig:phase2}
\end{figure}
Next, we elaborate on the regimes of the static conductivity. Taking a $\omega \rightarrow 0$ limit, we obtain the static conductivity in the two theoretical regimes as
\begin{align}
\rho_{xx}(T) = \begin{cases}
\dfrac{4 \pi^2 \mu}{Q_0^2}  & \mbox{for   } \gamma T \ll \mu(T),\\ \\
\dfrac{12\pi^2 \gamma^2 T^2}{Q_0^2 \mu} & \mbox{for   } \gamma T \gg \mu(T).
\end{cases} 
\label{eq:SMstatic1}
\end{align}

The bosonic mass-renormalization is evaluated in Eq.~(\ref{eqNb1}). Let us define $\tilde{g}_b=g_b/(4\pi)$. Next, we find the temperature scale $T^{\prime}_1$, where $\gamma T^\prime_1 = \mu_0+\tilde{g}_b T^{\prime}_1 \log(\gamma T^{\prime}_1/ \mu_0)$. Solving for $T^{\prime}_1$, we get
\begin{align}
T^\prime_1=-\frac{\mu_0}{\tilde{g}_b W[-\gamma/\tilde{g}_b\exp(-\gamma/\tilde{g}_b)]},
\label{eqn:Tscale}
\end{align}
where $W[x]$ is the Lambert $W$ function. For different coupling strength $\tilde{g}_b$, the form for this function is given by
\begin{align}
W\left[-\frac{\gamma}{\tilde{g}_b}\exp\left(-\frac{\gamma}{\tilde{g}_b}\right)\right]= \begin{cases}
-\dfrac{\gamma}{\tilde{g}_b} & \mbox{for   } \tilde{g}_b \geq \gamma,\\ \\
-\dfrac{\gamma}{\tilde{g}_b} \exp \left(-\frac{\gamma}{\tilde{g}_b}\right) & \mbox{for   } \tilde{g}_b < \gamma.
\end{cases} 
\label{eq:Lambert}
\end{align}
Putting this in Eq.~(\ref{eqn:Tscale}), we get the temperature scale
\begin{align}
T^\prime_1=\begin{cases}
\dfrac{\mu_0}{\gamma} & \mbox{for   } \tilde{g}_b \geq \gamma,\\ \\
\dfrac{\mu_0}{\gamma} \exp\left(\frac{\gamma}{\tilde{g}_b} \right) & \mbox{for   } \tilde{g}_b < \gamma.
\end{cases} 
\end{align}
It can be seen that if $\tilde{g}_b\geq\gamma$, the temperatures scale $T^{\prime}_1$ collapses on $\mu_0/\gamma$. Consequently, we are always in the $\gamma T<\mu(T)$ regime. In other words, ${\gamma T>\mu(T)}$ regime is never attained if the coupling between the bosons is stronger than that of the Landau damping coefficient. In main text Fig.~(1c), we have already presented the phase diagram for this scenario. In this regime, the static conductivity is given by
\begin{align}
\rho_{xx}(T) = \begin{cases}
\dfrac{4 \pi^2 \mu_0}{Q_0^2} +\dfrac{ 4 \pi^2 \tilde{g}_b }{Q_0^2}T \log\left(\frac{\gamma T}{\mu_0}\right) & \mbox{for   } \gamma T \gg \mu_0,\\ \\
\dfrac{4 \pi^2 \mu_0}{Q_0^2} & \mbox{for   } \gamma T \ll \mu_0.
\end{cases} 
\label{eq:SMstatic}
\end{align}
So, the incoherent charged bosons have linear-in-$T$ resistivity when $\gamma T \geq \mu_0$. This contribution also leads to the Drude form of optical conductivity, as shown in Eq.~(\ref{sigma_ana}). The bosonic contribution becomes independent of temperature below this temperature. However, the presence of conduction electrons will lead to a quadratic $T$-dependence of resistivity, just like in the Fermi liquid. 

Next, we focus on the situation when the interaction between the bosons is lower than the Landau damping constant, i.e., $\tilde{g}_b<\gamma$. In this situation, there will be an intermediate temperature regime, $\mu_0/\gamma < T < T^{\prime}_1$, where $\gamma T> \mu(T)$.  The resulting phase diagram is presented in Fig.~(\ref{fig:phase2}). The region bounded by the dotted line can harbor a non-Drude like optical conductivity as evaluated in Eq.~(\ref{sigma0_1}). The static conductivity in this limit is given by
\begin{align}
\rho_{xx}(T) \approx \begin{cases}  \dfrac{12\pi^2 \gamma^2  }{Q_0^2 \tilde{g}_b \log(\gamma T/\mu_0)}T & \mbox{for   } \gamma T \gg \mu_0,\\ \\
 \dfrac{48\pi^3 \gamma^2  }{Q_0^2 \mu_0}T^2 & \mbox{for   } \gamma T \ll \mu_0.
\end{cases} 
\label{eq:SMstatic3}
\end{align} 
Consequently, up to logarithmic corrections, we still have a linear-in-$T$ resistivity even when the bosonic interaction strength is weaker than the damping and $\gamma T > \mu_0$. However, such linear-in-$T$ resistivity does not subscribe to the Drude form of the optical conductivity. Below this temperature, the incoherent bosons also contribute to the $T^2$-resistivity expected in the Fermi-liquid regime. Thus, for weak coupling, the crossover from the strange metallic to Fermi-liquid behavior occurs through this intermediary region.

In the pseudogap phase, the opening of a gap at the temperature $T^{*}$ results
from a deconfining transition of a PDW order parameter into a SC and CDW fields. Above $T^{*}$, the incoherent bosons have a bare mass of $2\mu_0$. This is illustrated in more details in the next section. Using the bare mass for the bosons, a temperature region $T^{*}<T<T^{\prime}_0$ exists where the non-Drude form of the optical condutivity survives for weakly coupled bosons, i.e., $\tilde{g}_b<\gamma$.

\subsection{Bosonic bare mass in the ordered side}
\label{App:BareOrdered}
Near the ordered phase, i.e., just above $T^*$ in Fig.~(\ref{fig:phase2}), the bosonic propagator attains the bare mass due to the ordered parameter fluctuations. 
The Ginzburg-Landau free energy functional is given by
\begin{align}
\mathcal{F}[\psi]=\int d^d x\left[  \mu_0 \vert \psi(\mathbf{x}) \vert^2 + \frac{b}{2} \vert \psi(\mathbf{x}) \vert^4 \right].
\label{eqn:GLF}
\end{align}
If $\psi_0(\mathbf{x})$ minimizes $\mathcal{F}[\psi]$, we obtain
\begin{align}
\psi_0=\sqrt{-\frac{\mu_0}{b}}.
\end{align}
Expanding around the minima ${\psi(\mathbf{x})=\psi_0(\mathbf{x})+\delta \psi(\mathbf{x})}$, where ${\delta \psi(\mathbf{x})}$ is the fluctuations. Putting this in Eq.~(\ref{eqn:GLF}) and noting that the terms linear in $\delta \psi(\mathbf{x})$ vanishes, we obtain
\begin{align}
\mathcal{F}[\delta \psi]=\int d^d x\left[ -2 \mu_0 \vert \delta \psi(\mathbf{x}) \vert^2  + ... \right].
\end{align}
Therefore, the bare mass of the diffusive bosons just above the $T^*$ is given by $2\mu_0$.

\section{Mode-Mode Coupling: Higher order terms in self-energy}
\label{App:Pi2}
The second-order bosonic self-energy diagram -- which renormalizes both the mass-term $\mu$, and the imaginary term of the bosonic propagator, $\gamma$ --  is denoted by $\Pi_2(q_0)$ where $q_0$ is the external frequency. We emphasize that the finite-momentum bosons is dominant around $Q_0$, which is different from the external frequency in this diagram, $q_0$. The integral is given by
\begin{align}
\Pi_2(q_0)=g_b^2 \frac{1}{L^2} \sum_{\mathbf{k},\mathbf{p}} T^2 \sum_{\omega_n, \nu_n} &\mathcal{D}(\nu_n-\omega_n+ q_0,\mathbf{k}-\mathbf{p}) \nonumber \\ &\times \mathcal{D}(\nu_n,\mathbf{k})\mathcal{D}(\omega_n,\mathbf{p}).
\end{align}
Performing the summation over $\nu_n$ and $\omega_n$ and using the spectral decomposition, one readily obtains
\begin{widetext}
\begin{align}
\Pi_2(q_0)=g_b^2 \frac{1}{L^2} \sum_{\mathbf{k},\mathbf{p}} \int^{\infty}_{-\infty} \frac{dE_1}{2 \pi} \int^{\infty}_{-\infty} \frac{dE_2}{2 \pi} \int^{\infty}_{-\infty} \frac{dE_3}{2 \pi}
 &\left[ \mathcal{A}(E_1,\mathbf{a}) \mathcal{A}(E_2,\mathbf{b}) \mathcal{A}(E_3,\mathbf{d}) \left( n_B(E_2) - n_B(E_1)\right) \right]  \nonumber \\
 &\times  \left(\frac{n_B(E_3)-n_B(E_2-E_1)}{iq_0 -E_1+E_2-E_3}\right),
\end{align}
where we have defined
\begin{align}
\mathbf{a}&=\mathbf{k}^2+ \mu. \\
\mathbf{b}&=(\mathbf{k}-\mathbf{p})^2+\mu.\\
\mathbf{d}&= \mathbf{p}^2+ \mu.
\end{align}
Analytically continuing $i q_0 \rightarrow q_0 + i 0^+$, the imaginary part of the $\Pi_2$ becomes
\begin{align}
\text{Im } \Pi_2(q_0)=\frac{- g_b^2 q_0}{8 \pi^2} \frac{1}{L^2} \sum_{\mathbf{k},\mathbf{p}} \int^{\infty}_{-\infty} dE_1 \int^{\infty}_{-\infty} dE_2 \left[ \mathcal{A}(E_1,\mathbf{a}) \mathcal{A}(E_2,\mathbf{b}) \mathcal{A}(E_2-E_1+q_0,\mathbf{d}) \left( n_B(E_2) - n_B(E_1)\right) \right]\frac{\partial n_B}{\partial (E_2-E_1)}.
\end{align}

In the regime where $\vert E_2 -E_1 \vert <T$ and expanding the spectral function in the $q_0 \rightarrow 0$ limit, we obtain
\begin{align}
\text{Im } \Pi_2(q_0)=\frac{ \gamma g_b^2 T q_0}{ \pi^2} \frac{1}{L^2} \sum_{\mathbf{k},\mathbf{p}} \int^{\infty}_{-\infty} dE_1 \int^{\infty}_{-\infty} dE_2 \left[ \frac{\gamma E_1 \gamma E_2}{((\gamma E_2)^2 + \mathbf{a}^2)((\gamma E_1)^2 + \mathbf{b}^2)((\gamma (E_2-E_1))^2 + \mathbf{d}^2)} \left( \frac{n_B(E_2) - n_B(E_1)}{E_2-E_1}\right) \right].
\end{align}
Next, approximating $n_B(E)$ by using Eq.~(\ref{apprxnb}), the integrand will only contribute only both $\vert E_1 \vert <T $ and $\vert E_2 \vert <T$. Making a change of variables from $\tilde{E}=\gamma E$, we obtain

\begin{align}
\text{Im } \Pi_2(q_0)=\frac{ \gamma g_b^2 T q_0}{ \pi^2} \frac{1}{L^2} \sum_{\mathbf{k},\mathbf{p}} \int^{\gamma T}_{-\gamma T} d \tilde{E}_1 \int^{\gamma T}_{-\gamma T} d \tilde{E}_2 \frac{1}{(\tilde{E}^2_2+\mathbf{a}^2)(\tilde{E}^2_1+\mathbf{b}^2)((\tilde{E}_2-\tilde{E}_1)^2+\mathbf{d}^2)}.
\end{align}
\end{widetext}
Evaluating the integral in  the familiar regimes  $\gamma T \gg \mu$ and $\gamma T \ll \mu$, we obtain the forms
\begin{align}
\text{Im } \Pi_2(q_0) = \begin{cases} \dfrac{\gamma g_b^2 T^2 q_0}{ 2} \frac{1}{L^2} \sum_{\mathbf{k},\mathbf{p}} \frac{1}{\mathbf{a b d}(\mathbf{a}+\mathbf{b}+\mathbf{d})}
 & \mbox{for   } \gamma T \gg \mu, \\ \\
\dfrac{2 \gamma^3 g_b^2 T^4 q_0}{ \pi^2} \frac{1}{L^2} \sum_{\mathbf{k},\mathbf{p}} \frac{1}{\mathbf{a}^2\mathbf{b}^2 \mathbf{d}^2} & \mbox{for   } \gamma T \ll \mu.
\end{cases} 
\end{align}
Performing the momentum summation and arrive at expressions for the imaginary part $\Pi_2$ 

\begin{align}
\text{Im } \Pi_2(q_0) = \begin{cases} \dfrac{ c_1 \gamma g_b^2 T^2 }{ 16 \pi^3  \mu^2} q_0
 & \mbox{for   } \gamma T \gg \mu, \\ \\
\dfrac{ c_2 \gamma^3 g_b^2 T^4}{ 4 \pi^5  \mu^4}  q_0 & \mbox{for   } \gamma T \ll \mu,
\end{cases} 
\end{align}
where $c_1=0.323$ and $c_2=0.284$, which are evaluated numerically. On the other hand, the real part of $\Pi_2$ can be evaluated by utilizing Kramers-Kronig relations. The external frequency is taken to be small in the above calculations. Thus, a frequency cut-off $\lambda=\text{min }\left[ \mu, \gamma T\right]$ is used in the Kramers-Kronig relation. The Kramers-Kronig relation is given by
\begin{equation}
\text{Re }\Pi_2(q_0)=\frac{2}{\pi} \pv{\int_0^{\lambda}} \frac{\omega \text{Im }\Pi_2(\omega)}{\omega^2-q_0^2} d\omega.
\end{equation}
Therefore, the real-part of the $\Pi_2$ becomes

\begin{align} 
\text{Re }\Pi_2(q_0) = \begin{cases} \dfrac{ c_1 \gamma g_b^2 T^2}{8 \pi^4  \mu^2} \left( \lambda - q_0 \tanh^{-1} \left( \frac{\lambda}{q_0} \right) \right)
 & \mbox{for   } \gamma T \gg \mu, \\ \\
\dfrac{ c_2 \gamma^3 g_b^2 T^4}{ 2 \pi^6  \mu^4} \left( \lambda - q_0 \tanh^{-1} \left( \frac{\lambda}{q_0} \right) \right)  & \mbox{for   } \gamma T \ll \mu,
\end{cases} 
\end{align}
where $\lambda$ is the cut-off energy scale. Now evaluating the renormalization of the $\mu$ and $\gamma$ up to the second order for $\gamma T \gg \mu_0$, we get
\begin{eqnarray}
\mu &\approx& \mu_0 + \frac{g_b}{4 \pi} \log\left(\frac{\gamma T}{\mu_0}\right)+ \frac{2 c_1 \gamma \lambda }{\pi^2 \log^2\left( \gamma T/\mu_0 \right)} \\
\tilde{\gamma}&\approx&\gamma+\frac{c_1 \gamma }{ \pi \log^2\left( \gamma T/\mu_0 \right)}.
\end{eqnarray}
Taking the limit $\gamma T/\mu_0 \gg 1$, it is clear that the second-order terms are negligible. Next, evaluating the same for $\gamma T \ll \mu_0$, we get
\begin{eqnarray}
\mu &\approx& \mu_0 + \frac{c_2 \lambda (\gamma T)^4 }{2 \pi^6 \gamma  \mu_0^4}, \\
\tilde{\gamma}&\approx&\gamma+\frac{c_2 (\gamma T)^3}{4 \pi^5  \gamma \mu_0^4}.
\end{eqnarray}
Again, taking the limit $\gamma T/\mu_0 \ll 1$, it becomes clear that the higher-order terms are negligible compared to the first order ones. 

\subsection{Fate of bosonic vertex corrections}
\label{App:Vertex}
\begin{figure}[h!]
\includegraphics[width=0.45\textwidth]{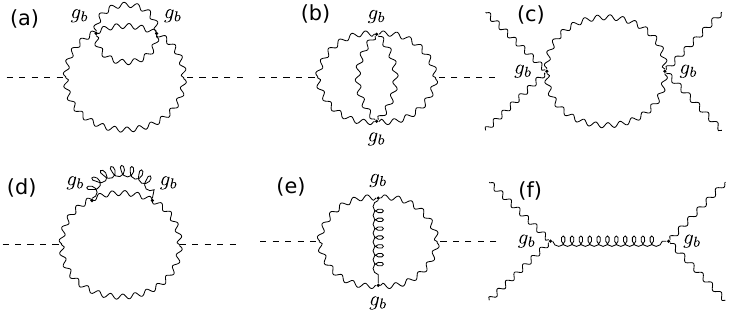}
\caption[0.5\textwidth]{\textbf{(a)} Shows the diagram associated with the current-current
correlation function with second order self-energy corrections. \textbf{(b)} Exhibits the same with bosonic vertex corrections. \textbf{(c)} Shows the bosonic bubble that can be replaced with the diagram shown in \textbf{(f)}. \textbf{(d-e)} Again shows the current-current correlation function by replacing the bosonic bubble with the curly line.} 
\label{fig:fdvert}
\end{figure}
Here, we discuss the bosonic vertex correction diagram at second-order in $g_b$ and compare it with the self-energy diagram given by $\Pi_2$ evaluated in the previous section. The second-order self energy diagram is shown in Fig.~(\ref{fig:fdvert}a), whereas the vertex correction diagram is presented in Fig.~(\ref{fig:fdvert}b). Next, we replace the boson-boson bubble in Fig.~(\ref{fig:fdvert}c) using the curly-line composite propagator, as shown in Fig.~(\ref{fig:fdvert}f). After this, it can be readily seen that the diagrams in Fig.(\ref{fig:fdvert}a) and Fig.~(\ref{fig:fdvert}b) can be replaced by those in Fig.~(\ref{fig:fdvert}d) and Fig.(\ref{fig:fdvert}e), respectively. Subsequently, we define the bosonic composite propagator represented by the curly-line in Fig.~(\ref{fig:fdvert} f) as $\mathcal{F}(\mathbf{q},\omega)$. In this notation, the diagram in Fig.~(\ref{fig:fdvert}d) becomes
\begin{align}
\Xi_1= \frac{T^2}{L^2} \sum_{\mathbf{k,q},\omega_n,\Omega_n} \mathcal{D}^3(\mathbf{k},\omega) \mathcal{D}(\mathbf{k+q},\omega+\Omega) \mathcal{F}(\mathbf{q},\Omega).
\end{align}
Similarly, the diagram in Fig.~(\ref{fig:fdvert}(e)) becomes
\begin{align}
\Xi_2= \frac{T^2}{L^2} \sum_{\mathbf{k,q},\omega,\Omega} \mathcal{D}^2(\mathbf{k},\omega) \mathcal{D}^2(\mathbf{k+q},\omega+\Omega) \mathcal{F}(\mathbf{q},\Omega).
\end{align}
Now, after performing analytical continuation, we obtain using the retarded form of the bosonic Green's function
\begin{align}
T \sum_{\omega_n} \mathcal{D}(\mathbf{k},\omega) &= \int_{-\infty}^{\infty} \frac{d\omega}{2 \pi} \frac{1}{-i \omega + k^2 + \mu} \nonumber \\
&= i \int_{-\infty}^{\infty}  \frac{d l_\tau}{2 \pi} \frac{1}{ l_\tau + k^2 + \mu}, 
\end{align}
where in the last step we have made a simple change of variable. From the last definition, it is clear that 
\begin{align}
\frac{\partial\mathcal{D}(\mathbf{k+q},\omega+\Omega)}{\partial (-i \omega)}  =-\mathcal{D}^2(\mathbf{k+q},\omega+\Omega).
\end{align}
Next, using these previous relations to evaluate $\Xi_2$, we get
\begin{align}
\Xi_2= -\frac{i T}{L^2} \sum_{\mathbf{k,q},\Omega} \int_{-\infty}^{\infty}  \frac{d l_\tau}{2 \pi} \mathcal{D}^2(\mathbf{k},l_\tau) \partial_{l_\tau}\mathcal{D}(\mathbf{k+q},l_\tau+\Omega) \mathcal{F}(\mathbf{q},\Omega).
\end{align}
Integrating by parts, we find
\begin{align}
\Xi_2= -2\frac{i T}{L^2} \sum_{\mathbf{k,q},\Omega} \int_{-\infty}^{\infty}  \frac{d l_\tau}{2 \pi} \mathcal{D}^3(\mathbf{k},l_\tau) \mathcal{D}(\mathbf{k+q},l_\tau+\Omega) \mathcal{F}(\mathbf{q},\Omega).
\end{align}
Finally, reverting back to the earlier notation, we get
\begin{align}
\Xi_2 &= -2\frac{T^2}{L^2} \sum_{\mathbf{k,q},\omega_n,\Omega_n} \mathcal{D}^3(\mathbf{k},\omega) \mathcal{D}(\mathbf{k+q},\omega+\Omega) \mathcal{F}(\mathbf{q},\Omega) \nonumber \\ 
  &=-2\Xi_1.
\end{align}
We have already argued in the previous section that the $\Xi_1$-correction due to $\Pi_2$ is negligible in all regimes. Since we find that $\Xi_2$ if of the same order of magnitude as $\Xi_1$, the vertex correction diagram in Fig.~(\ref{fig:fdvert} b) can also be safely ignored in our analysis.  

\section{Hall conductivity}
\label{App:Hall}
To discuss the effect of magnetic field, in the first order in magnetic field, we calculate the Hall conductivity which is given by
\begin{align}
\sigma_{xy}^{\left(1\right)} = \frac{i H}{\omega_{n}}T\sum_{\varepsilon_{n}} \frac{1}{L} &\sum_{\mathbf{q}}\left[ q_{x} \mathcal{D}({\varepsilon_{n},\mathbf{q}})\partial_{q_{x}}\mathcal{D}({\varepsilon_{n}+\omega_{n},\mathbf{q}}) \right. \nonumber \\ &\left.
-q_{y}\mathcal{D}({\varepsilon_{n},\mathbf{q}})\partial_{q_{y}}\mathcal{D}({\varepsilon_{n}+\omega_{n},\mathbf{q}}) \right].
\label{Hall}
\end{align}
For a particle-hole symmetric theory, the Hall conductivity is naturally expected to vanish. This means that the incoherent bosons at finite-$\mathbf{Q}$ do not contribute to the Hall conductivity. Using the fact that $\partial_{q_{x}} \mathcal{D}(x)=q_x \mathcal{D}^2(x)$, only the wave-vector near $q_x=Q_0$ will contribute. As a result, we obtain
\begin{align}
\sigma_{xy}^{\left(1\right)}(\omega_n)  =\frac{i H Q_0^2} {\omega_{n}} T\sum_{\varepsilon_{n}} \frac{1}{L}\sum_{\mathbf{q}} &\left[ \mathcal{D}(\varepsilon_n,\mathbf{q}) \mathcal{D}^2(\varepsilon_n+ \omega_n, \mathbf{q}) \right. \nonumber \\ &\left. - \mathcal{D}^2(\varepsilon_n, \mathbf{q}) \mathcal{D}(\varepsilon_n + \omega_n,\mathbf{q}) \right].
\end{align}
Performing the Matsubara summation by using spectral functions, we arrive at
\begin{align}
\sigma_{xy}^{\left(1\right)}(\omega)  =\frac{i H Q_0^2}{L} &\sum_{\mathbf{q}} \int^{\infty}_{-\infty} \frac{dE_1}{2 \pi} \int^{\infty}_{-\infty} \frac{dE_2}{2 \pi} \frac{n_B(E_1)-n_B(E_2)}{\omega(E_1-E_2+\omega)} \nonumber \\
&\times \left( \mathcal{A}(E_1,\mathbf{q}) \tilde{\mathcal{A}}(E_2,\mathbf{q})- \tilde{\mathcal{A}}(E_1,\mathbf{q})\mathcal{A}(E_2,\mathbf{q}) \right),
\end{align}
where $\mathcal{A}(E1,\mathbf{q})$ is given in Eq.~(\ref{Spectral1}) and the $\tilde{\mathcal{A}}(E1,\mathbf{q})$ is given by
\begin{equation}
\tilde{\mathcal{A}}(q,E)=-2 \text{Im }[\mathcal{D}_R^2(E,\mathbf{q})]=-\frac{4 \gamma E ( q^2 + \mu)}{(\gamma E)^2+( q^2+\mu)^2}.
\label{Asq}
\end{equation}
Therefore, taking the $\omega \rightarrow 0$, the expression for the Hall conductivity becomes
\begin{align}
\sigma_{xy}^{\left(1\right)}(0)  =\frac{i H Q_0^2}{L}  \sum_{\mathbf{q}} \int^{\infty}_{-\infty} \frac{dE_1}{2 \pi} &\int^{\infty}_{-\infty} \frac{dE_2}{2 \pi} \mathcal{A}(E_1,\mathbf{q}) \tilde{\mathcal{A}}(E_2,\mathbf{q}) \nonumber \\
 &\times \left[ \frac{\coth(\frac{E_1}{2T})-\coth(\frac{E_2}{2T})}{(E_1-E_2)^2} \right].
\end{align}
This can be trivially shown to be exactly zero by noting that the $\mathcal{A}(E,\mathbf{q})$, $\tilde{\mathcal{A}}(E,\mathbf{q})$ and $\coth(E)$ are all anti-symmetric functions with respect to $E$. Since $I(-E_1,-E_2)=-I(E_1,E_2)$, as a consequence, the incoherent bosons will indeed have a vanishing Hall conductivity.

\section{Second Moment of Conductivity}
\label{App:Sigma2}
The second moment of the conductivity, the term proportional to the square of the field $H$, is given in terms of the bosonic Green's function by
\begin{align}
\sigma^{(2)}_{xx}(\omega_n)=-\frac{H^2}{\omega_n} \text{Im } T\sum_{\varepsilon_n} \frac{1}{L} &\sum_{\mathbf{q}} \partial_{q_y} \mathcal{D}(\varepsilon_n,\mathbf{q}) \nonumber \\ &\times \partial_{q_y}\mathcal{D}(\varepsilon_n + \omega_n,\mathbf{q}).
\end{align}
Using the form of bosonic propagator $\mathcal{D}(\omega_n,\mathbf{q})$, we obtain
\begin{align}
\sigma^{(2)}_{xx}(\omega_n)=-\frac{4  Q_0^2 H^2}{\omega_n} \text{Im } &\left\lbrace T\sum_{\varepsilon_n} \frac{1}{L} \sum_{\mathbf{q}} \mathcal{D}^2(\varepsilon_n,\mathbf{q})
 \right. \nonumber \\ &\left. \times  \mathcal{D}^2(\varepsilon_n + \omega_n,\mathbf{q}) \right\rbrace. 
\end{align}
The spectral function in Eq.~(\ref{Asq}) is used to perform the Matsubara summation over $\varepsilon_n$. After analytical continuation, the real part of the second moment of conductivity becomes
\begin{align}
\sigma^{(2)}_{xx}(\omega_n)=-\frac{4  Q_0^2 H^2}{\omega_n L} \sum_{\mathbf{q}} \int_{-\infty}^{\infty} dE_1 &\tilde{\mathcal{A}}(E_1,\mathbf{q})  \tilde{\mathcal{A}}(E_1+\omega,\mathbf{q}) \nonumber \\  &\times \frac{\partial n_B}{\partial E_1}.
\end{align}
The Bose function is approximated by Eq.~(\ref{apprxnb}) and the momentum summation is carried out by replacing $(q^2+\mu)=t$, i.e.,
\begin{equation}
\sigma^{(2)}_{xx}(\omega \rightarrow 0)=-\frac{4 T Q_0^2 H^2}{\pi^2} \int_{\mu}^{\infty} t^2 dt\int_{-\infty}^{\infty} dE_1 \frac{\gamma^2}{\left\lbrace  (\gamma E_1)^2 + t^2 \right\rbrace}.
\end{equation}
Finally, performing the integral over $E_1$ and $t$, and then by expanding in the two familiar limits, we obtain the expression for the real part of static second moment of conductivity
\begin{align}
\sigma^{(2)}_{xx}= \begin{cases} 
\dfrac{ 8 \gamma^2 Q_0^2 T^2 H^2}{ 5 \pi^2 \mu^5}  & \mbox{for   } \gamma T \ll \mu,\\ \\
\dfrac{5 T \gamma Q_0^2  H^2}{ 16 \pi \mu^4} 
 & \mbox{for   } \gamma T \gg \mu.
\label{Apsigma2}
\end{cases} 
\end{align}

\section{Polarization bubble due to the Zeeman field}
\subsection{Singlet Case}
\label{App:Singlet}
For particle-particle pairs of singlets, the contribution to the self-energy due to the Zeeman term is evaluated here. The correction to the mass term is given by
\begin{align}
\Pi(H,\mathbf{Q}_0)=\frac{g_I^2}{L} \sum_k T \sum_{\varepsilon_n} &\left[ \mathcal{G}(-\varepsilon_n,\xi_{-\mathbf{k},\uparrow})\mathcal{G}(\varepsilon_n,\xi_{\mathbf{k}+\mathbf{Q}_0,\downarrow}) \nonumber \right. \\  & \left. - \mathcal{G}(-\varepsilon_n,\xi_{-\mathbf{k},\downarrow})\mathcal{G}(\varepsilon_n,\xi_{\mathbf{k}+\mathbf{Q}_0,\uparrow}) \right],
\label{singletGF}
\end{align} 
where $\xi_{k,\sigma}=k^2-\sigma H$ where $\sigma=\pm 1$. Next, performing the Matsubara summation over $\varepsilon_n$, we arrive at the expression
 which is independent of magnetic field. The mass term thus becomes
\begin{equation}
\mu=\mu_0+\mu_T,
\end{equation}
where $\mu_T=\tilde{g}_b T \log(\gamma T/\mu_0)$. So the mass-term has no contribution from the Zeeman field.

\subsection{Triplet Case}
\label{App:Triplet}
Here, we calculate the self-energy correction due to the bosons formed with paired electrons of triplet spin-symmetry. The corresponding expression is given by
\begin{equation}
\Pi(H,\mathbf{Q}_0)=\frac{g_I^2}{L} \sum_{\mathbf{k}} T \sum_{\varepsilon_n} \mathcal{G}(-\varepsilon_n,\xi_{-\mathbf{k},\uparrow})\mathcal{G}(\varepsilon_n,\xi_{\mathbf{k}+\mathbf{Q}_0,\uparrow}),
\end{equation} 
where $\xi_{k,\sigma}=k^2-\sigma H$ where $\sigma=\pm 1$, in our units $\hbar^2/(2m_e)=1$. Performing the $\varepsilon_n$-summation, we get
\begin{equation}
\Pi(H,\mathbf{Q}_0)=\frac{g_I^2}{L} \sum_{\mathbf{k}} \left\lbrace \frac{1-n_F(\xi_{\mathbf{k}}-H)-n_F(\xi_{\mathbf{k}+\mathbf{Q}_0}-H)}{\xi_{\mathbf{k}+\mathbf{Q}_0}+\xi_{\mathbf{k}}-2H} \right\rbrace.
\end{equation}
Next, using a flat band approximation, we can write the momentum summation in the following form
\begin{align}
\Pi(H,\mathbf{Q}_0)=\frac{\mathcal{N}(\epsilon_F)g_I^2}{4 \pi^2} &  \int_0^{2\pi} d\theta \int_{0}^{\Lambda} d\xi \nonumber \\ &\times \frac{\tanh(\frac{\xi+\zeta-H}{2T})+\tanh(\frac{\xi-H}{2T})}{2 \xi +\zeta-2H},
\label{eq:triplet_main}
\end{align}
where $\Lambda$ is the largest energy scale of the system. Additionally, we have substituted $\zeta \equiv Q_0^2 + 2 k_F Q_0 \cos(\theta)$. Now at $T\rightarrow 0$, we will use that $\tanh(x/T)\rightarrow \text{sgn }(x)$ and then performing the $\xi$-integral we arrive at
\begin{widetext}
\begin{align}
\Pi(H,\mathbf{Q}_0) = \frac{\mathcal{N}(\epsilon_F)g_I^2}{4 \pi^2} \int_0^{2\pi} d\theta \begin{cases}
\log \left(1-\frac{2H}{\zeta} \right)+\log \left(-\dfrac{\zeta+2\Lambda-2H}{\zeta}\right) & \mbox{for   } \zeta \leq 0,\\ \\
\log \left(\dfrac{\zeta+2\Lambda-2H}{\zeta} \right) & \mbox{for   } \zeta > 0 \text{ and  } H-\zeta \leq 0.
\end{cases} 
\label{eq:Apptr8}
\end{align}
\end{widetext}
Recall that the $\Lambda$ is the ultraviolet energy cutoff, and hence, expanding in $H \ll 2 k_F Q_0 \ll \Lambda$, we get
\begin{align}
\Pi(H,\mathbf{Q}_0) = C -\frac{\mathcal{N}(\epsilon_F)g_I^2}{4 \pi^2} \int^{2\pi-p_1} _{p_1}  d\theta \frac{2 H} {Q_0^2+2 k_F \cos(\theta)},
\end{align}
where $p_1\equiv \cos^{-1}(Q_0/(2 k_F))$ and $C$ is the $H$-independent constant. Now integrating over $\theta$, one obtains
\begin{align}
\Pi(H,\mathbf{Q}_0) = C + \frac{2 \gamma}{\pi} \coth^{-1}\left(\frac{2k_F+Q_0}{\sqrt{4 k_F^2 - Q_0^2}}\right) H, 
\end{align}
where we have used the definition of $\gamma$ from Appendix~(\ref{App:Scattering}). 

The constants can be absorbed in the bare bosonic mass, $\mu_0$. Therefore, the total bosonic mass renormalization due to the Zeeman field $H$, becomes 
\begin{align}
\mu =  
 \mu_0 + \mu_T+ \alpha H,
\label{eq:muH}
\end{align}
where $\alpha\equiv\frac{2 \gamma}{\pi} \coth^{-1}\left(\frac{2k_F+Q_0}{\sqrt{4 k_F^2 - Q_0^2}}\right)$ and we also define $\mu_H \equiv \alpha H$. Thus, we obtain the mass-renormalization due to the Zeeman field, which is used to evaluate magnetoresistance in the next section.

\subsection{Magnetoresistance}
\label{App:MR}
In this section, we explicitly show the calculations to arrive at the magnetoresistance for diffusive bosons. The magnetoresistance quantifies the change of resistance due to the application of the magnetic field and is given by
\begin{equation}
\frac{\Delta \rho_{xx} (H)}{\rho_{xx}(H=0)}=\frac{\rho_{xx} (H)-\rho_{xx} (0)}{\rho_{xx} (0)}.
\end{equation}

The complete resistivity tensor in terms of the conductivity is written as~\cite{murayama2008mesoscopic}
\begin{equation}
\rho_{xx}=\frac{\sigma_{xx}}{\sigma^2_{xx}+\sigma^2_{xy}}.
\end{equation}
Notice that, for incoherent transport, we have shown in Appendix~\ref{App:Hall} that $\sigma_{xy}=0$ and hence the expression for the magnetoresistance in terms of conductivity simply reads
\begin{equation}
\frac{\Delta \rho_{xx} (H)}{\rho_{xx}(H=0)}=\frac{\sigma_{xx}(0)-\sigma_{xx}(H)}{\sigma_{xx}(H)}.
\label{MR_1}
\end{equation}
Next, the expression for $\sigma_{xx}=\sigma^{(0)}_{xx}+\sigma^{(2)}_{xx}$ where $\sigma^{(0)}_{xx}$ is already calculated in Eq.~(\ref{sigma_ana}) and Eq.~(\ref{sigma0_1}) and $\sigma^{(2)}_{xx}$ is evaluated in Eq.~(\ref{Apsigma2}).

\subsubsection{Singlet Case}
\label{App:MR_Singlet}
Here, the renormalization of the mass term is independent of the magnetic field and is given by $\mu=\mu_0+\mu_T$. The regimes are given by the maximum of $\mu_0$ and $\mu_T$. So the expression for the magnetoresistance becomes
\begin{equation}
\frac{\Delta \rho_{xx} (H)}{\rho_{xx}(0)}=\frac{\kappa H^2}{\beta + \kappa H^2},
\end{equation}
where $\beta$ and $\kappa$ are the coefficients of $H$ in $\sigma^{(0)}_{xx}$ and $\sigma^{(2)}_{xx}$, respectively.
We consider that the interaction between the bosons is larger than the Landau damping coefficient, i.e., ${\tilde{g}_b>\gamma}$. In this scenario, if we take the limit $\gamma T/\mu \ll 1$ in Eq.~(\ref{Apsigma2}) and Eq.~(\ref{sigma_ana}), it becomes clear that $\sigma^{(2)}_{xx}$ is negligible compared to the $\sigma^{(0)}_{xx}$. Hence, we have the leading contribution to the MR by taking the limit $\kappa/\beta \ll 1$  
\begin{equation}
\frac{\Delta \rho_{xx} (H)}{\rho_{xx}(0)}\approx \frac{\kappa}{\beta} H^2,
\end{equation}
where in the first regime when $\mu_0>\mu_T$ the constant is given by $\frac{\kappa}{\beta}\equiv-\frac{ 32 \gamma^2 T^2}{ 5 \mu_0^4}$. By contrast, when $\mu_0<\mu_T$, the constant is given by $\frac{\kappa}{\beta}\equiv-\frac{ 32 \gamma^2 T^2}{ 5 (\mu_0+\mu_T)^4}$. Thus, the bosons arising from the singlet pairing of electrons have the same dependence on $H$ as the conduction electrons would do in the typical Fermi liquid.  

When the interaction between the bosons is weaker than the Landau damping expanding in $\kappa/\beta \gg 1$, the magnetoresistance becomes independent of $H$ for singlet particle-particle pairs.

\subsubsection{Triplet Case}
\label{App:MR_Triplet}
\begin{figure}[h!]
\includegraphics[width=0.45\textwidth]{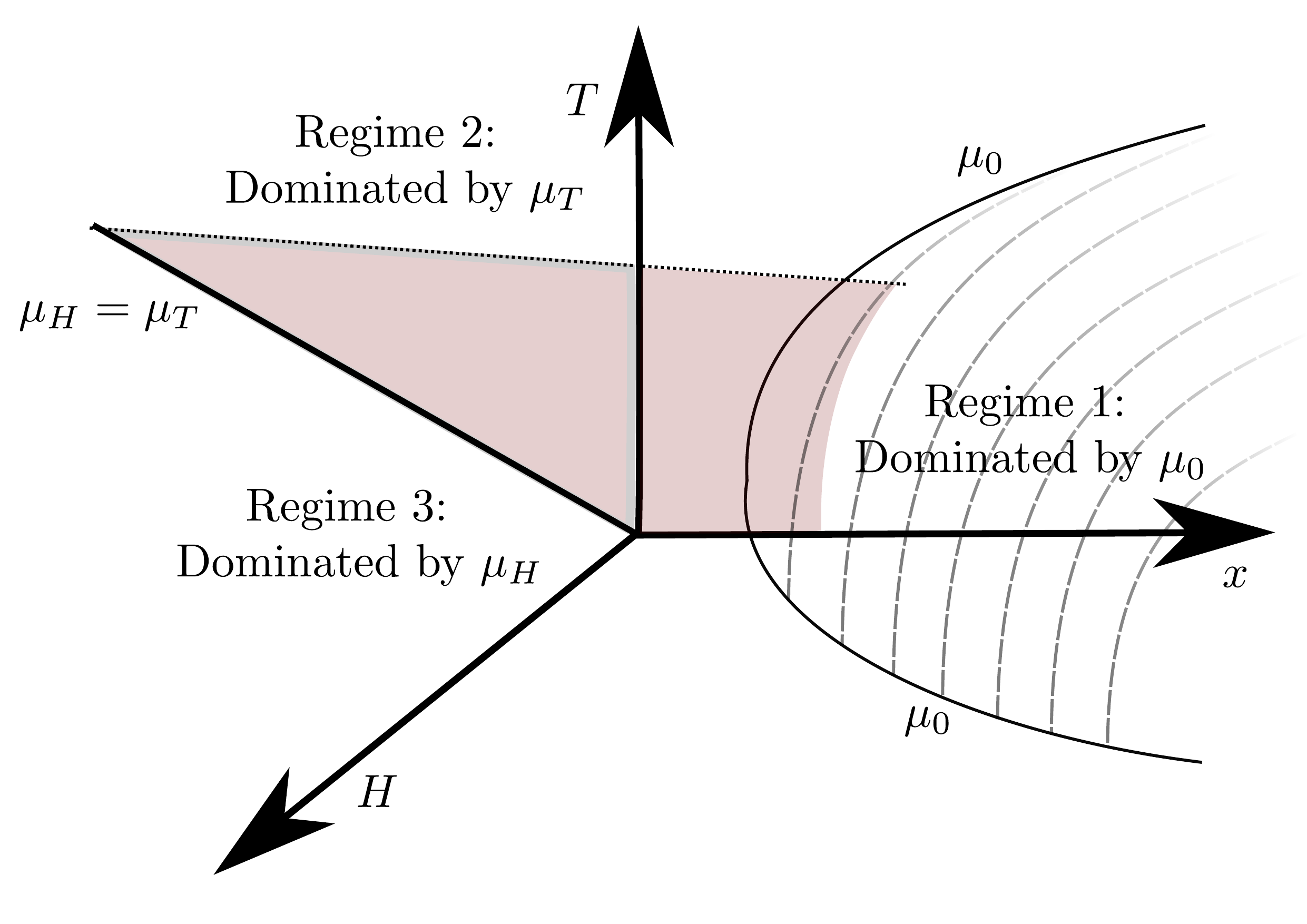}
\caption[0.75\textwidth]{The figure illustrates the different regimes in the temperature, doping and magnetic field plane. The mass term renormalization for the particle-particle pairs is given by $\mu=\mu_0+\mu_T+\mu_H$. The maximum of the three mass scales determines the regime: In regime 1, the mass is dominated by $\mu_0$; similarly, in regime 2 and 3, it is dominated by $\mu_T$ and $\mu_H$, respectively.}
\label{figreg}
\end{figure}
Next, we perform the calculation of the renormalization of the mass term when the bosons emerge from pairs of
high-energy electrons that have spin-triplet symmetry. The bosonic mass correction due to the Zeeman field is evaluated in Eq.~(\ref{eq:muH}). Similarly, the expressions for $\sigma^{(0)}_{xx}$ in terms of $\mu$ are evaluated in Eq.~(\ref{eq:SMstatic1}) and the same for $\sigma^{(2)}_{xx}$ are evaluated in Eq.~(\ref{Apsigma2}). Notice we have different regimes depending on the renormalization of the mass term from bosonic interactions and the Zeeman field. These regimes are illustrated in Fig.~(\ref{figreg}) in the magnetic field, hole doping and temperature plane. The different scenarios arise because the mass term is either dominated $\mu_0$, $\mu_T$ or $\mu_H$. We elaborate on the different possibilities one by one in the following. 

\subsubsection{$\tilde{g}_b\geq\gamma$ and $\mu_T \ll \mu_H$}
First, if the interaction between the bosons is larger than the Landau damping coefficient, i.e., ${\tilde{g}_b>\gamma}$, we are always in ${\gamma T \ll \mu}$. Additionally, if we are in a regime dominated by the magnetic field scale , i.e., $\mu_T\ll \mu_H $ (see regime 3 in Fig.~(\ref{figreg})), the mass correction coming from the Zeeman field is given by ${\mu=\mu_0+\mu_T+\alpha H}$ in Eq.~(\ref{eq:muH}). Therefore, the magnetoresistance evaluates to
\begin{equation}
\frac{\Delta \rho_{xx} (H)}{\rho_{xx}(0)}=\dfrac{\dfrac{Q_0^2}{4 \pi^2 (\mu_0+\mu_T)}-\dfrac{Q_0^2}{4 \pi^2 (\mu_0+\mu_T+\alpha H)}-\sigma^{(2)}_{xx}(H)}{\dfrac{Q_0^2}{4 \pi^2 (\mu_0+\mu_T+\alpha H)}+\sigma^{(2)}_{xx}(H)}.
\end{equation}
If we take the limit $\gamma T/\mu \ll 1$ in Eq.~(\ref{Apsigma2}), it is clear that the $\sigma^{(2)}_{xx}$ becomes negligible. Therefore, the equation for MR becomes
\begin{align}
\frac{\Delta \rho_{xx} (H)}{\rho_{xx}(0)} &\approx \dfrac{\dfrac{1}{\mu_0+\mu_T}-\dfrac{1}{ (\mu_0+\mu_T+\alpha H)}}{\dfrac{1}{\mu_0+\mu_T+\alpha H}}, \nonumber \\
\frac{\Delta \rho_{xx} (H)}{\rho_{xx}(0)} &= \frac{\alpha}{\mu_0+\mu_T} H.
\end{align}
Therefore, we obtain a linear-in-$H$ magnetoresistance in the regime 3 of Fig.~(\ref{figreg}). Note that $\mu_H \gg \mu_T$ can be interpreted as $H \gg \eta T$, where $\eta=\frac{\mu_0+\tilde{g}_b \log(\gamma T/\mu_0)}{\alpha}$. Thus up to logarithmic corrections $\eta$ is just a constant. We emphasize that this a similar high-field regime where linear-in-$H$ magnetoresistance is observed~\cite{Hussey2020}. 

\subsubsection{$\tilde{g}_b\geq\gamma$ and $\mu_T\gg \mu_H$}
Second, we still keep the interaction between the bosons stronger than the Landau damping coefficient, i.e., ${\tilde{g}_b>\gamma}$. However, if the temperature-correction is larger than the magnetic field scale, i.e., $\mu_T\gg \mu_H$, the mass correction coming from the Zeeman field is independent of the field and is given by $\mu=\mu_0+\mu_T$ (see regime 2 in Fig.~(\ref{figreg})). Consequently, the evaluation of magnetoresistance becomes similar to the one performed for the singlet in Appendix~(\ref{App:MR_Singlet})
\begin{equation}
\frac{\Delta \rho_{xx} (H)}{\rho_{xx}(0)}\approx \frac{\kappa}{\beta} H^2,
\end{equation}
where $\frac{\kappa}{\beta}\equiv-\frac{ 32 \gamma^2 T^2}{ 5 (\mu_0+\mu_T)^4}$. Again for $\mu_H \ll \mu_T$ can be written as $H \ll \eta T$. Therefore, in the low-field regime shows a quadratic $H$-dependence of magnetoresistance. 

\subsubsection{$\tilde{g}_b\geq\gamma$ for $\mu_T\ll \mu_0$ and $\mu_H\ll \mu_0$}
Similarly, if the temperature or field correction of the bosonic mass term is smaller than the bare bosonic mass, i.e., $\mu_T\ll \mu_0$ and $\mu_H \ll \mu_0$, the mass correction coming from the Zeeman field is independent of the field and is given by $\mu=\mu_0$ (see regime 1 in Fig.~(\ref{figreg})). Again, the magnetoresistance becomes
\begin{equation}
\frac{\Delta \rho_{xx} (H)}{\rho_{xx}(0)}\approx \frac{\kappa}{\beta} H^2,
\end{equation}
here we get $\frac{\kappa}{\beta}\equiv-\frac{ 32 \gamma^2 T^2}{ 5 \mu_0^4}$. So again we have a $H^2$-dependence of magnetoresistance in the regime 1 of Fig.~(\ref{figreg}). In this regime we have already established the conventional Fermi liquid behavior. 

Therefore, when the interaction between the bosons is stronger than the Landau damping coefficient the MR is given by
\begin{align}
\dfrac{\Delta \rho_{xx} (H)}{\rho_{xx}(0)} =  \begin{cases}
\dfrac{\kappa}{\beta} H^2 & \mbox{in   } \text{regimes 1 and 2},\\ \\
\dfrac{\alpha}{\mu_0+\mu_T} H & \mbox{in   } \text{regime 3},
\end{cases} 
\label{eq:MR_app}
\end{align}
where the coefficient $\kappa/\beta$ is different in regimes 1 and 2. Notice that such an $H$-evolution of magnetoresistance is recently observed in overdoped cuprates~\cite{Hussey2020}.

\subsubsection{$\tilde{g}_b<\gamma$}
When the coupling is weaker than the Landau damping, a temperature regime survives where $\mu\ll \gamma T$ (for details, refer to Appendix~ \ref{App:Static}). We demand the limit $\mu/(\gamma T) \ll 1$ and recognize that $\sigma^{(0)}_{xx}$ is negligible. Consequently, using the expression of $\sigma^{(2)}_{xx}$ from Eq.~(\ref{Apsigma2}) in the expression of MR in Eq.~(\ref{MR_1}). We notice that the MR becomes independent of $H$ in all the temperature regimes for $\tilde{g}_b<\gamma$.

\subsection{On the quadrature form of the magnetoresistance}
\label{App:ChangeinMR}
This section provides more details in order to compare the scaling of the in-plane magnetoresistance with that observed experimentally. The in-plane MR is given by
\begin{equation}
\Delta \rho_{xx} = \rho_{xx}(H,T)-\rho_{xx}(0,0)=\frac{1}{\sigma_{xx}(H,T)}-\frac{1}{\sigma_{xx}(0,0)},
\label{Quad1}
\end{equation}
where in the second equality we have used the fact that the Hall conductivity vanishes. Near the QCP, $\Delta \rho_{xx}$ experimentally displays a quadrature dependence~\cite{hayes2016scaling,Hayes18,Hussey2020} as follows
\begin{align}
\Delta \rho_{xx} &= \sqrt{a^2 T^2 + b^2 H^2},
\label{sds5}
\end{align}
where $a$ and $b$ are constants. As we explained in the main text, in the low-field and high-field limits, this quantity scales as
\begin{align}
\Delta \rho_{xx} \propto  \begin{cases}
H & \mbox{for   } H \gg T ,\\ \\
\dfrac{H^2}{T} & \mbox{for }   H \ll T.
\end{cases} 
\label{eq:MR_app_12}
\end{align}
Although our phenomenological model cannot determine exactly the quadrature dependence of Eq.~(\ref{sds5}),
our results for the scaling behavior in both low-field and high-field limits can suggest a similar quadrature ansatz. 
We concentrate on the physical regime
 when the interaction between the bosons is stronger than the Landau damping parameter, i.e.,
 $\tilde{g}_b \ge \gamma$, i.e., $\mu \gg \gamma T$. We also restrict our attention to the case when the bosons
 emerge from pairs of high-energy electrons that have spin-triplet symmetry. Consequently, the mass-term is given by Eq.~(\ref{eq:muH}).
 The maximum among $\mu_0$, $\mu_T$, and $\mu_H$ determines the regime, 
as shown in Fig.~(\ref{figreg}). Let us first focus on regime 3 of Fig.~(\ref{figreg}), where the mass term
 is dominated by $\mu_H$. Mathematically, we are in the regime $\mu_H \gg \mu_T \gg \mu_0$, or $H \gg \eta T$, 
where up to logarithmic corrections,  $\eta$ is only a constant. Using the form of $\sigma^{(0)}_{xx}$ 
from Eq.~(\ref{eq:SMstatic1}) and $\sigma^{(2)}_{xx}$ from Eq.~(\ref{Apsigma2}) in Eq.~(\ref{Quad1}), we get
\begin{align}
\Delta \rho_{xx} = \frac{1}{\frac{Q_0^2}{4 \pi^2 (\mu_0+ \mu_T+\mu_H)}+\sigma^{(2)}_{xx}}-\frac{4 \pi^2 \mu_0}{Q_0^2}.
\label{sds}
\end{align} 
Since the interaction between the bosons is stronger than the Landau damping parameter, i.e., $\tilde{g}_b \ge \gamma$, by taking $(\gamma T)/\mu \ll 1$ in Eq.~(\ref{Apsigma2}), $\sigma^{(2)}_{xx} \rightarrow 0$. Consequently, we get
\begin{align}
\Delta \rho_{xx} \approx \frac{4 \pi^2}{Q_0^2} \left( \mu_T +\mu_H \right).
\label{sds1}
\end{align}
Therefore, in the high field regime $H \gg \eta T$ (i.e., the regime 3 of Fig.~(\ref{figreg})), the leading order $H$-dependence is given by
\begin{align}
\Delta \rho_{xx} \propto H.
\label{asd1}
\end{align} 
The next regime is when the mass-term is dominated by $\mu_T$ (i.e., the regime 2 in Fig.~(\ref{figreg})). Notice that this is the low-field regime, $H \ll \eta T$. Here, we have
\begin{align}
\Delta \rho_{xx} = \frac{1}{\frac{Q_0^2}{4 \pi^2 (\mu_0+ \mu_T)}+\frac{ 8 \gamma^2 Q_0^2 T^2 H^2}{ 5 \pi^2 (\mu_0+\mu_T)^5}}-\frac{4 \pi^2 \mu_0}{Q_0^2}.
\label{sds2}
\end{align}
However, we cannot ignore $\sigma^{(2)}_{xx}$ to get the leading order $H$-dependence, since $\sigma^{(0)}_{xx}$ is independent of the field. Expanding in powers of $H$, we obtain

\begin{align}
\Delta \rho_{xx} = \frac{4 \pi^2 \mu_T }{Q_0^2} - \frac{128 \pi^2}{5 Q_0^2 \tilde{g}_b^3 \log((\gamma T)/\mu_0)^3} \frac{H^2}{T}.
\label{sds3}
\end{align}
Therefore, in the low-field regime $H \ll \eta T$, the leading order scaling is given by
\begin{align}
\Delta \rho_{xx} \propto \frac{H^2}{T}.
\label{sds34}
\end{align}
Next, in the regime 1 of Fig.~(\ref{figreg}), the mass-term is dominated by $\mu_0$. In the latter regime, the in-plane magnetoresistance is given by
\begin{align}
\Delta \rho_{xx} = \frac{4 \pi^2 \mu_T }{Q_0^2} - \frac{128 \pi^2}{5 Q_0^2 \tilde{g}_b^3 \log((\gamma T)/\mu_0)^3} \frac{H^2}{T}.
\label{sds33}
\end{align}
Therefore, in the Fermi liquid regime, the leading order scaling is given by
\begin{align}
\Delta \rho_{xx} \propto \frac{H^2 T^2}{\mu_0^3}.
\label{sds41}
\end{align}
Finally, combining Eq.~(\ref{asd1}), Eq.~(\ref{sds34}), and Eq.~(\ref{sds41}), we have the scaling of $\Delta \rho_{xx}$, to leading order in $H$, as
\begin{align}
\Delta \rho_{xx} \propto  \begin{cases}
H & \mbox{for   } \text{regime 3},\\ \\
\dfrac{H^2}{T} & \mbox{for }   \text{regime 2},\\ \\
\dfrac{H^2 T^2}{\mu_0^3} & \mbox{for }   \text{regime 1},
\end{cases} 
\label{eq:MR_app_11}
\end{align}
which is identical to the scaling observed from the quadrature dependence in Eq.~(\ref{eq:MR_app_12}). We conclude that, although our calculations cannot determine exactly the quadrature dependence of $\Delta \rho_{xx}$ presented in Eq.~(\ref{sds5}), we can find a similar scaling behavior n the low-field and high-field limits.
\bibliography{Cuprates.bib}
\end{document}